\pdfoutput=1 
\documentclass[11pt]{article}

\usepackage{jheppub}
\usepackage{mathtools}        
\usepackage{amssymb, amsthm}
\usepackage{bm}                 
\usepackage{physics}            
\usepackage{tensor}            
\DeclareRobustCommand{\ts}{\tensor}
\usepackage{cancel}            
\usepackage{environ}            
\usepackage{xcolor}
\usepackage{graphicx}
\usepackage{tikz-cd}
\usepackage{float}
\usepackage{orcidlink}

\providecommand{\m}{}\renewcommand{\m}{\mu}
\providecommand{\n}{}\renewcommand{\n}{\nu}
\providecommand{\r}{}\renewcommand{\r}{\rho}
\providecommand{\s}{}\renewcommand{\s}{\sigma}
\providecommand{\la}{}\renewcommand{\la}{\lambda}

\providecommand{\g}{}\renewcommand{\g}{\gamma}
\providecommand{\om}{}\renewcommand{\om}{\omega}

\newcommand{\cE}{\mathcal{E}}     
\newcommand{\bth}{\Theta}         
\newcommand{\Om}{\Omega}          
\newcommand{\La}{\Lambda}         
\newcommand{\p}{\partial}         
\newcommand{\Ga}{\Gamma}          

\providecommand{\sb}{}\renewcommand{\sb}[1]{\left( #1 \right)}

\newcommand{\TensorWithLabel}[3]{%
    \vphantom{\overset{(#1)}{#2}}%
    \ts{%
        \smash[t]{\overset{\mathclap{\scalebox{0.5}{$(#1)$}}}{#2}}%
        \vphantom{#2}%
    }{#3}%
}

\newcommand{\DefineTensorCommands}[2]{%
    \expandafter\providecommand\csname t#1\endcsname{}%
    \expandafter\renewcommand\csname t#1\endcsname[1]{\ts{#2}{##1}}%
    \expandafter\providecommand\csname ct#1\endcsname{}%
    \expandafter\renewcommand\csname ct#1\endcsname[1]{\TensorWithLabel{\mathrm{C}}{#2}{##1}}%
    \expandafter\providecommand\csname gt#1\endcsname{}%
    \expandafter\renewcommand\csname gt#1\endcsname[1]{\TensorWithLabel{\tilde{\mathrm{G}}}{#2}{##1}}%
    \expandafter\providecommand\csname mft#1\endcsname{}%
    \expandafter\renewcommand\csname mft#1\endcsname[1]{\TensorWithLabel{-4}{#2}{##1}}%
    \expandafter\providecommand\csname mtt#1\endcsname{}%
    \expandafter\renewcommand\csname mtt#1\endcsname[1]{\TensorWithLabel{-2}{#2}{##1}}%
    \expandafter\providecommand\csname zt#1\endcsname{}%
    \expandafter\renewcommand\csname zt#1\endcsname[1]{\TensorWithLabel{0}{#2}{##1}}%
    \expandafter\providecommand\csname tt#1\endcsname{}%
    \expandafter\renewcommand\csname tt#1\endcsname[1]{\TensorWithLabel{2}{#2}{##1}}%
    \expandafter\providecommand\csname ft#1\endcsname{}%
    \expandafter\renewcommand\csname ft#1\endcsname[1]{\TensorWithLabel{4}{#2}{##1}}%
    \expandafter\providecommand\csname lct#1\endcsname{}%
    \expandafter\renewcommand\csname lct#1\endcsname[1]{\TensorWithLabel{\mathrm{LC}}{#2}{##1}}%
    \expandafter\providecommand\csname ggt#1\endcsname{}%
    \expandafter\renewcommand\csname ggt#1\endcsname[1]{\TensorWithLabel{\mathrm{G}}{#2}{##1}}%
}

\DefineTensorCommands{E}{E}       
\DefineTensorCommands{T}{T}       
\DefineTensorCommands{cE}{\cE}    
\DefineTensorCommands{bth}{\bth}  
\DefineTensorCommands{Om}{\Om}    
\DefineTensorCommands{om}{\om}
\DefineTensorCommands{La}{\La}    
\DefineTensorCommands{Ga}{\Ga}    
\DefineTensorCommands{na}{\nabla} 
\DefineTensorCommands{R}{R}       
\DefineTensorCommands{S}{S}       

\DefineTensorCommands{bE}{\bm{E}}
\DefineTensorCommands{bcE}{\bm{\cE}}
\DefineTensorCommands{bbth}{\bm{\Theta}}
\DefineTensorCommands{bOm}{\bm{\Omega}}

\newenvironment{eqs}
    {\begin{equation}\begin{aligned}}
    {\end{aligned}\end{equation}}

\NewEnviron{subeqs}{%
    \begin{subequations}
    \begin{align}
        \BODY
    \end{align}
    \end{subequations}
}

\newcommand{\XH}{X_{\textrm{\tiny H}}}
\newcommand{\XP}{X_{\textrm{\tiny P}}}

\newcommand{\eq}[2]{\begin{equation} #1 \label{#2} \end{equation}}

\newcommand{\vmakro}{\mathfrak{v}}

\begin{document}

\title{Galilean cosmologies in two dimensions}

\author{\orcidlink{0000-0001-7980-5394}Daniel Grumiller}
\author[1]{\orcidlink{0000-0003-1858-0405}and Kai-Yu Zhang\note{Corresponding author}}
\affiliation{Institute for Theoretical Physics, TU Wien, Wiedner Hauptstrasse 8, A-1040 Vienna, Austria}

\emailAdd{grumil@hep.itp.tuwien.ac.at, kaiyu.zhang@student.tuwien.ac.at}

\abstract{
Inspired by Carroll black holes, we consider the most general theory of Galilean dilaton gravity in two dimensions. We find generically that the solution space describes cosmological spacetimes. From a limiting perspective, these Galilean geometries connect the white and black hole singularities through the bifurcation surface. The latter turns into a Galilean extremal surface in the limit of infinite speed of light. We also discuss aspects of Galilean tachydynamics and find that the entropy is given by the dilaton evaluated at the Galilean extremal surface. 
}

\preprint{TUW--26--04}
\maketitle

\enlargethispage{1truecm*}

\section{Introduction}

Gravity in two dimensions (2d) provides an abundance of toy models for classical and quantum gravity to test ideas and concepts that can provide lessons for higher dimensions, see e.g.~\cite{Grumiller:2002nm,Mertens:2022irh} and Refs.~therein. One of these concepts is the consideration of degenerate signatures $(0,\dots,0,-,\dots,-,+,\dots,+)$ that has become popular due to numerous Carrollian applications in the past decade, see e.g.~\cite{Bagchi:2025vri,Ciambelli:2025unn,Nguyen:2025zhg,Ruzziconi:2026bix} and Refs.~therein. In 2d, up to reinterpretations and conventional choices, the only option that is degenerate but not completely degenerate is a signature $(0,+)$.

When the degenerate direction corresponds to time, the signature $(0,+)$ describes Carroll dilaton gravity \cite{Grumiller:2020elf,Gomis:2020wxp} which in turn features Carroll black holes as part of its solution space \cite{Ecker:2023uwm}. As suggested in that work, it might be of interest to search for the Galilean counterpart of Carroll black holes, where the roles of space and time are interchanged. This is the main goal of the present work.

The punchline of our paper is that most of the 2d results can be translated directly from Carrollian to Galilean dilaton gravity, as anticipated from the isomorphism of 2d Carroll and Galilei algebras, but the interpretation changes drastically since we are exchanging space and time. In particular, instead of black holes we find generically cosmologies as part of the state space and instead of a Hawking temperature they exhibit a characteristic velocity.

In our work, the term Galilean cosmology refers to a homogeneous Galilean spacetime equipped with an absolute clock and a time-dependent spatial metric. In contrast to conventional Lorentzian cosmology, the spatial metric is the non-degenerate part of a degenerate spacetime geometry. The cosmological evolution is therefore encoded in the time dependence of the spatial geometry rather than in a Lorentzian spacetime metric.

For related work in three-dimensional Galilean gravity and holography see, e.g., 
\cite{Papageorgiou:2009zc,Papageorgiou:2010ud,Bagchi:2010zz,Bagchi:2012xr,Bagchi:2014iea,Bergshoeff:2016lwr,Hartong:2016yrf,Bergshoeff:2017btm,Matulich:2019cdo,Merbis:2019wgk,Gomis:2019nih,Aviles:2026qjl}.

This paper is organized as follows. In Section \ref{sec:2}, we review the most general Galilean dilaton gravity model in 2d presented from various perspectives --- gauge theoretic, second order, and dimensional reduction. We also construct all classical solutions and derive a Galilean version of the Birkhoff theorem: generic solutions have a Killing vector associated with homogeneity of the cosmological solution. In Section \ref{sec:3}, we elaborate on Galilean cosmologies by defining first G-tachy manifolds and then discussing Galilean tachydynamics, i.e., the momentum response to changes in velocity and the first law connecting velocity, momentum, and entropy. Then we discuss the Galilean version of extremal surfaces and use them to define Galilean cosmologies. In Section \ref{sec:4}, we discuss the evolution and global structure of homogeneous Galilean cosmologies. In Section \ref{sec:5}, we present several examples, including Galilei--Schwarzschild--Tangherlini and Galilei--Jackiw--Teitelboim models. We summarize and conclude in the final Section. Appendix \ref{app:A} collects Galilean symmetry formulas. Appendix \ref{app:B} addresses aspects of Galilean gravity. Appendix \ref{app:C} reviews the large-$c$ expansion of Einstein gravity.

\section{Galilean dilaton gravity in two dimensions}\label{sec:2}

In this Section, we present and discuss Galilean dilaton gravity in 2d. In contrast to earlier work \cite{Grumiller:2020elf,Gomis:2020wxp,Aviles:2022xyx} our discussion is not based on the Bargmann or Newton--Hooke algebra but rather on the (non-extended) Galilei algebra, which we deform in this Section in a way that is consistent with Galilei boosts. Our framework closely parallels the 2d Carroll dilaton gravity construction in \cite{Grumiller:2020elf,Ecker:2023uwm}.

We start with the Poisson $\sigma$-model (PSM) formulation in Subsection \ref{sec:2.1} and continue with the second-order formulation in Subsection \ref{sec:2.2}. Subsection \ref{sec:2.3} considers Galilean dilaton gravity as a Galilean limit of Lorentzian dilaton gravity in 2d. In Subsections \ref{sec:2.4} and \ref{sec:2.5}, we construct all solutions for a general class of models. Finally, in Subsection \ref{sec:2.6}, we prove a Galilean version of the Birkhoff theorem, i.e., the existence of at least one Killing vector for all solutions.

\subsection{PSM formulation}\label{sec:2.1}

The Poisson Sigma Model (PSM) framework allows for a natural formulation of 2d dilaton gravity \cite{Ikeda:1993fh,Schaller:1994es}. The PSM is a topological field theory defined on a worldsheet $\Sigma_2$. Its scalar fields $X^I$ define a map from $\Sigma_2$ to an $N$-dimensional target space and are accompanied by 1-form gauge fields $A_I$. Its action takes the form
\begin{align}
    S[X^I, A_I] =\frac{k}{2\pi} \int_{\Sigma_2} \left( X^I\, \dd A_I + \frac{1}{2} P^{IJ}(X^K)\, A_I \wedge A_J \right)\label{psm action}
\end{align}
with indices $I,J,K = 1,\dots,N$. The Poisson tensor $P^{IJ}(X^K)$ satisfies the nonlinear Jacobi identities
$P^{IL}\p_L P^{JK} + \text{cyclic}(I,J,K) = 0$. Here, $\p_I \equiv \p/\p_{X_I}$ denotes differentiation with respect to the target space coordinates.

To obtain 2d Galilean gravity, we fix the target space dimension to $N=3$. We then identify the gauge fields $A_I$ and the scalars $X^I$ as
\begin{equation}
     A_I = (\omega, \tau, e) \qquad\qquad
     X^I = (X, \XH, \XP)\,.
\end{equation}
Here, $\tau$ and $e$ are the temporal and spatial einbeins, while $\omega$ denotes the Galilean spin connection. The scalar $X$ is the dilaton, while $\XH$ and $\XP$ are auxiliary fields that implement the torsion constraints.

The dynamical structure of the theory is encoded in the Poisson tensor $P^{IJ}$. To realize the Galilean algebra in 2d (see Appendices \ref{Galilean algebra} and \ref{app:B}), we choose the only non-vanishing components of the Poisson tensor to be $P^{12} = \XP$ and $P^{23} = \mathcal{V}(X^I)$, leading to
\begin{align}
    P^{IJ} &= \begin{pmatrix}
    0 & \XP & 0 \\
    -\XP & 0 & \mathcal{V}(X,\XP) \\
    0 & -\mathcal{V}(X,\XP) & 0
    \end{pmatrix}\,.\label{poisson tensor}
\end{align}
The Jacobi identity constrains the potential $\mathcal{V}$ to be independent of $\XH$. 

The only difference to Newton--Cartan dilaton gravity presented in \cite{Grumiller:2020elf} is that we do not have a fourth target space coordinate (called $X_{\textrm{\tiny M}}$ there and corresponding on-shell to the mass) nor the associated 1-form $m$.

Substituting the Poisson tensor \eqref{poisson tensor} into the action \eqref{psm action} gives the first-order action for Galilean dilaton gravity.
\begin{align}
    S = \frac{k}{2\pi}\,\int_{\Sigma_2} \Big( X R[G] + \XH R[H] + \XP R[P] + \mathcal{V}(X, \XP) \,\tau \wedge e \Big) \label{2d first order action}
\end{align}
The term $X R[G] = X \dd \omega$ describes the non-minimal coupling of the dilaton to the scalar curvature. The fields $\XH$ and $\XP$ serve as Lagrange multipliers that enforce constraints on the intrinsic torsion $R[H] = \dd \tau$ and on the spatial torsion $R[P] = \dd e + \omega \wedge \tau$, respectively. The final term, $\mathcal{V}(X,\XP)$, represents the potential density.

The action \eqref{2d first order action} is explicitly invariant under Galilei boost transformations and is equipped with a larger gauge symmetry generated by $\lambda_I = (\lambda_G, \lambda_H, \lambda_P)$. The corresponding transformations are given by 
\begin{equation}
\delta_{\lambda} A_{I} = \dd\lambda_{I} + (\p _{I} P^{LM}) A_{L} \lambda_{M} \qquad\qquad
\delta _{\lambda} X^{I}  = \lambda_{J} P^{JI}
\label{psm symm X}
\end{equation} 
The PSM gauge symmetries for the Galilean dilaton model act as 
\begin{subequations}
\begin{align}
    \delta \omega_\m &= \p_\m \lambda_G + (\p_X \mathcal{V}) \left( \lambda_P \tau_\m - \lambda_H e_\m \right) &  \delta X &= -\lambda_H \XP  \\
    \delta \tau_\m &= \p_\m \lambda_H & \delta \XH &= \lambda_G \XP - \lambda_P \mathcal{V} \\
    \delta e_\m &= \p_\m \lambda_P - \lambda_G \tau_\m + \lambda_H \omega_\m + (\p_P \mathcal{V}) \left( \lambda_P \tau_\m - \lambda_H e_\m \right) & \delta \XP &= \lambda_H \mathcal{V} \,.\label{gauge e}
\end{align}
\end{subequations}
The gauge parameter $\lambda_G$ generates Galilean boosts. The remaining two gauge symmetries can be interpreted on-shell as diffeomorphisms generated by a vector field $\xi^\mu$, using the standard map $\lambda_I=A_{I\,\mu}\,\xi^\mu$~\cite{Schaller:1994pa}.

In principle, the potential $\mathcal{V}(X,\XP)$ is arbitrary. In this work, we consider a typical class of potentials
\begin{align}
    \mathcal{V}(X, \XP) = V(X) + \frac{1}{2} U(X) \XP^2\label{def mathcal V}
\end{align}
where $V(X)$ denotes the dilaton self-interaction term, and $U(X)$ encodes the non-minimal coupling to the torsion sector. 

In Section~\ref{sec:2.3}, we demonstrate that the action \eqref{2d first order action} arises from the non-relativistic limit of 2d Lorentzian dilaton gravity.

Varying the action \eqref{2d first order action} with respect to all fields yields the equations of motion:
\begin{subeqs}
    \delta X & & \text{Galilean curvature:} & & R[G] &= \dd \omega = -\partial_{X} \mathcal{V} \, \tau \wedge e \label{eom:curvature} \\
    \delta \XH & & \text{No intrinsic torsion:} & & R[H] &= \dd \tau = 0 \label{eom:intrinsic_torsion} \\
    \delta \XP & & \text{Spatial torsion:} & & R[P] &= \dd e + \omega \wedge \tau = -\partial_{P} \mathcal{V} \, \tau \wedge e \label{eom:spatial_torsion} \\
    \delta \omega & & \text{Galilean clock one form:} & & \dd X + \XP \tau &= 0 \label{eom:dilaton} \\
    \delta \tau & & \text{Casimir conservation:} & & \dd \XH - \XP \omega + \mathcal{V} e &= 0 \label{eom:casimir} \\
    \delta e & & \text{Auxiliary evolution:} & & \dd \XP - \mathcal{V} \tau &= 0 \label{eom:auxiliary}
\end{subeqs}
The first equation \eqref{eom:curvature} specifies the Galilean curvature 2-form. It is generically non-vanishing, unless $\mathcal{V}$ is independent of the dilaton. The second equation \eqref{eom:intrinsic_torsion} enforces a vanishing intrinsic torsion, $\dd \tau = 0$, guaranteeing the presence of an absolute time, which is equivalent to the torsion-free condition in Newton--Cartan gravity~\cite{Hartong:2022lsy}. This constraint is automatically satisfied because the potential $\mathcal{V}$, as defined in Eq.~\eqref{def mathcal V}, does not depend on $\XH$, which is a direct consequence of the Jacobi identities.  The third equation \eqref{eom:spatial_torsion} links the spatial torsion to the derivative of the potential with respect to $\XP$; for a generic potential, this yields a non-vanishing spatial torsion proportional to $\XP U(X)$. The fourth equation \eqref{eom:dilaton} imposes a geometric constraint on the dilaton which leads to $\theta^\mu \partial_\mu X = 0$, where $\theta$ is the Galilei-invariant spatial vector defined uniquely through \eqref{eq:def theta} which satisfies $\theta^\mu e_{\mu}=1, \theta^\mu \tau_{\mu}=0$, implying that the dilaton field must be spatially homogeneous. The fifth equation \eqref{eom:casimir} entails the conservation of the Casimir function. Lastly, the sixth equation \eqref{eom:auxiliary} determines the temporal evolution of the auxiliary field $\XP$ along the absolute time direction.

Consider a Weyl rescaling of the gauge fields together with a transformation of the scalar fields and the spin connection. The equations of motion \eqref{eom:curvature}--\eqref{eom:auxiliary} remain invariant under the following set of transformations, parameterized by a scalar function $Q(X)$
\begin{subeqs}
    e &\to e^\prime = e^{\frac{1}{2}Q(X)} e &\quad \XP &\to \XP^\prime = e^{-\frac{1}{2}Q(X)} \XP \\
    \tau &\to \tau^\prime = e^{\frac{1}{2}Q(X)} \tau &\quad  \XH &\to \XH^\prime = e^{-\frac{1}{2}Q(X)} \XH \\
    \omega &\to \omega^\prime = \omega +  \frac{1}{2}\p_XQ(X) (\XH \tau + \XP e) &\quad X &\to X^\prime = X
    \label{Weyl map}
\end{subeqs}
upon changing the original potential to a shifted one,
\begin{align}
    \mathcal{V} \to \mathcal{V}^\prime = e^{-Q(X)} \left[ \mathcal{V} +\frac{1}{2} \XP^2\p_X Q(X) \right]\,. 
\end{align}

This relation between different models shows that the solution space of a more generic model with a potential $\mathcal{V}$ given by \eqref{def mathcal V} can be mapped to that of a simpler model with potential depending only on the dilaton by a suitable choice of $\frac{1}{2}Q(X)$ by choosing
\begin{equation}
    Q(X) =  \int^{X} U(y) \dd y\,. \label{def Q}
\end{equation}
Thus, one can drop the non-minimal coupling term $U(X)$ to obtain the solution space of the associated $U=0$ model with potential $\bar{\mathcal{V}} = \bar{V}(X)$, and finally map it back to the solution space of the original model with non-zero $U$.

\subsection{Second-order action}
\label{sec:2.2}

In this Subsection, we derive the second-order formulation of the theory by integrating out a certain part of the boost connection $\omega$ and the auxiliary scalar $\XP$. This procedure introduces a new Lagrange multiplier, denoted by $\r$, which encodes the undetermined sector of spin connection and cannot be fixed uniquely. The resulting non-uniqueness of the spin connection in non-Lorentzian gravity is a well-known feature; see, for instance, \cite{Duval:2014uoa,Hartong:2015xda,Andringa:2010it}.

The analysis starts by splitting the Galilean boost connection $\omega$ into a torsionless part $\hat{\omega}$ and a contorsion 1-form $\tilde{\omega}$.
\eq{
    \omega = \hat{\omega} + \tilde{\omega}
}{omega decomp}
The torsion-free connection $\hat{\omega}$ is defined as the solution of the spatial torsion constraint $R[P] = 0$. 
\begin{equation}
    \dd e + \hat{\omega} \wedge \tau = 0
\end{equation}
In the 2d setting, this relation fixes the spatial part of $\hat{\omega}$. We then require that $\hat{\omega}$ has no component along the temporal direction, and all contributions in the temporal direction are absorbed into the second term $\tilde{\omega}$, which yields
\eq{
    \hat{\omega}_\m = v^\n(\dd e)_{\m\n} = K e_{\m}
}{omegahat}
where $K$ denotes the extrinsic curvature, defined as 
\eq{
    K=\theta^\m \theta^\n K_{\m\n} \qquad\qquad K_{\m\n}=-\frac{1}{2}\mathcal{L}_{v}\sb{e_\m e_\n}
}{eq:K}
where $v^\mu$ is the temporal vector defined through \eqref{eq:def theta} and it is not Galilei-invariant, see the gauge transformation \eqref{eq:delta v}. All residual freedom in the boost connection is encoded in the 1-form $\tilde{\omega}$. This field is determined by the spatial torsion equation of motion, which now includes the effect from the generally non-constant potential $\mathcal{V}$.
\begin{align}
 \dd e + \omega \wedge \tau = -\p_P\mathcal{V}\, \tau \wedge e
\end{align}
Inserting the decomposition \eqref{omega decomp} and employing the relation $\partial_{P}\mathcal{V} = U(X)\XP$ yields
\begin{align}
    \tilde{\omega} = \rho\, \tau + U(X)\, \XP\, e
\end{align}
where the scalar field $\rho$ serves as a Lagrange multiplier. Combining these expressions, the complete connection can be written as
\eq{
    \omega_\m = \hat{\omega}_\m + U(X)\, \XP\, e_\m + \rho\, \tau_\m\,.
}{general omega}

We now determine the auxiliary scalar field $\XP$. Contracting both sides of the Galilean clock 1-form equation \eqref{eom:dilaton} 
with the inverse temporal einbein $v^\mu$ yields
\eq{
    \XP = v^\mu \p_\mu X\,. 
}{XP=vpX}

We now insert the solution for $\omega$ in Eq.~\eqref{general omega} together with the expression~\eqref{XP=vpX} for $\XP$ into the first-order Lagrangian \eqref{2d first order action}. After integrating the term $X \dd \hat{\omega}$ by parts we use the geometric relations
\begin{align}
    \dd \hat{\omega} &= -v^\m v^\n \hat{R}_{\m\n}\,\mathrm{vol} &
    \dd \tau &= -\theta^\mu a_\mu \,\mathrm{vol} &
    \mathrm{vol} &:= \tau \wedge e = \det(\tau,e)\,\dd^2 x
\end{align}
where $\hat{R}_{\m\n}$ denotes the Galilean Ricci tensor, associated with the minimal and torsionless spin connection $\hat{\om}$ defined in Appendix \ref{app:B}, cf.~Eq.~\eqref{2Ricci}, and $a_\mu = \mathcal{L}_v \tau_\mu$ represents the acceleration vector. The kinetic contribution is simplified  by
\begin{align}
    \dd X \wedge \tau 
    = -(\theta^\mu \p_\mu X)\,\mathrm{vol}\,.
\end{align}
Finally, the resulting second-order Lagrangian density
\begin{align}
    \mathcal{L}_{\text{2nd}} 
    = \det(\tau,e)\big( 
          V(X)
        + \tfrac{1}{2} U(X) (v^\mu \p_\mu X)^2-v^\m v^\n \hat{R}_{\m\n} X
        + \rho\, (\theta^\mu \p_\mu X)
        - \XH (\theta^\mu a_\mu)  
    \big)
    \label{2nd action}
\end{align}
is a functional of the geometric data $(\tau,e)$, the dilaton $X$, and two Lagrange multipliers. The first one, $\rho$, imposes the constraint $\theta^\mu \p_\mu X = 0$, thereby enforcing spatial homogeneity of the dilaton. The second one, $\XH$, imposes the vanishing of the intrinsic torsion component 
\begin{equation}
\theta^\mu a_\mu = v^\m \theta^\n (\dd \tau)_{\m\n}=0\,.
\end{equation}

Under a Galilean boost with parameter $\lambda$, the fields transform according to $\delta_\lambda X=0,\delta_\lambda \omega = \dd \lambda$, $\delta_\lambda \tau = 0$, and $\delta_\lambda e = -\lambda \tau$. The transformation behavior of $\rho$ is fixed by demanding consistency with the decomposition in Eq.~\eqref{general omega}.
Solving this relation for $\delta_\lambda \rho$ yields
\begin{align}
    \delta_\lambda \rho 
    = -\hat{\nabla}_\mu (\lambda v^\mu) +\lambda U(X)\sb{v^\mu\p_\mu X} 
\end{align}
where $\hat{\nabla}$ is the covariant derivative associated with $\hat{\omega}$ via the vielbein postulate. Finally, the other Lagrange multiplier transforms under Galilean boosts as
\eq{
    \delta_\lambda \XH=\lambda\sb{v^\mu \p_\mu X}\,.
}{eq:whatever}
With this transformation law, $\delta_\lambda v^\mu=-\lambda\,\theta^\mu$, and noticing the scalar $\theta^\m a_\m$ is Galilei invariant, the second-order action $S_{\mathrm{2nd}}=\frac{k}{2\pi}\int\mathcal{L}_{\mathrm{2nd}}$
is invariant under Galilean boosts. 

\subsection{Galilean dilaton gravity as a limit}
\label{sec:2.3}

\begin{figure}[h!tb]
\centering
\begin{tikzcd}[column sep=3.7cm, row sep=2.0cm]
\parbox{4.5cm}{\centering
$D$-dimensional Einstein gravity}
\arrow[r, "\substack{\text{spherical}\\ \text{reduction}}"]
\arrow[d, "\substack{\text{large-}c\\ \text{limit}}"']
&
\parbox{4.5cm}{\centering
2d spherical-reduction\\
dilaton gravity}
\arrow[d, "\substack{\text{large-}c\\ \text{limit}}"']
\\
\parbox{4.5cm}{\centering
$D$-dimensional Galilean gravity}
\arrow[r, "\substack{\text{spherical}\\ \text{reduction}}"']
&
\parbox{4.5cm}{\centering
2d Galilean dilaton gravity\\
\eqref{2nd action}}
\end{tikzcd}
\caption{Spherical reduction and the Galilean large-$c$ limit. The two routes lead to the same 2d Galilean dilaton gravity bulk action, modulo boundary terms.}
\label{derivation_path}
\end{figure}
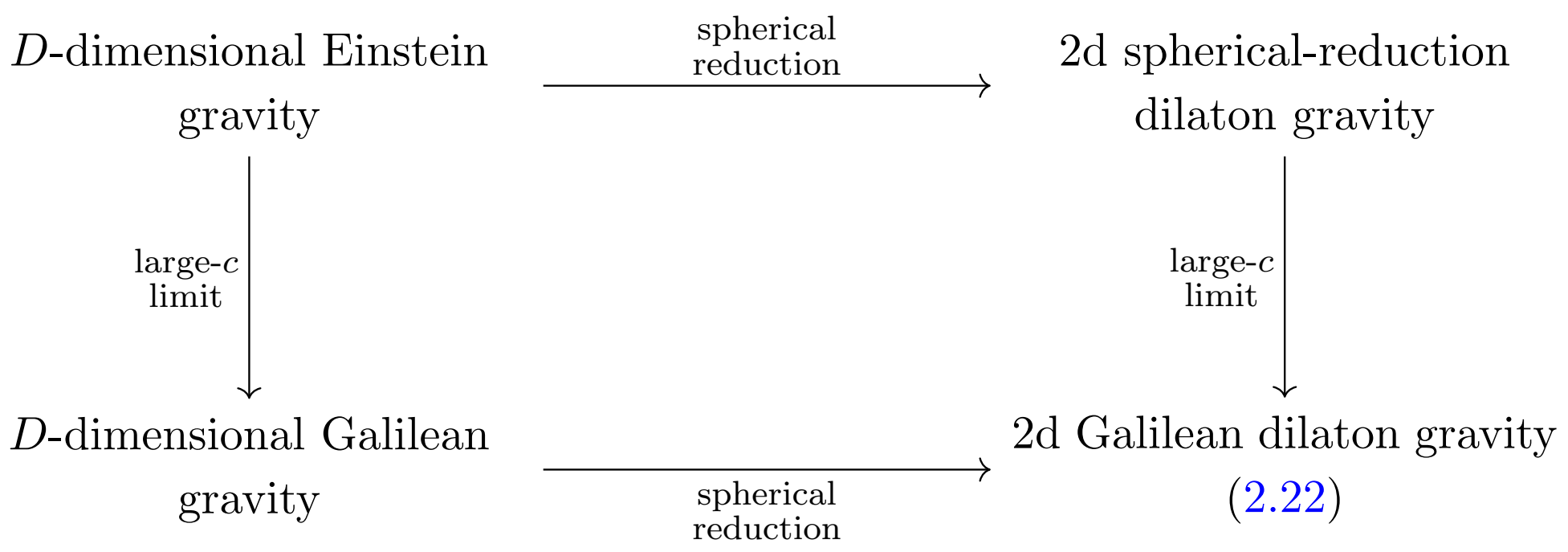

This Subsection establishes two complementary derivations of the Galilean dilaton theory from relativistic theories. We first take the large-$c$ limit directly in a generic two-dimensional (2d) Lorentzian dilaton gravity model with quartic potential and show that it reproduces the second-order Galilean action \eqref{2nd action}. We then start from $D$-dimensional Einstein gravity and perform the large-$c$ expansion first before imposing spherical symmetry. On the spherical ADM form considered below and suitable $c$-dependent rescales of $G_D$ and radial function $\mathcal{R}$, the resulting 2d bulk action agrees with the second-order Galilean action \eqref{2nd action} again. Hence in this sense, non-relativist limit and  dimensional reduction commute, as shown in Fig~\ref{derivation_path}.

\subsubsection{Two-dimensional Galilean limit}
The action of a general 2d Lorentzian dilaton gravity theory with a term quadratic in the dilaton gradient (see Section 2.1.3 of \cite{Grumiller:2002nm} for a review) 
\begin{align}
    I_{\text{dil}}=\frac{c^3k_{\mathrm{rel}}}{4\pi}\int\dd^2x\sqrt{-g}\left[XR+2V(X)-(\p X)^2 U(X)\right]\label{relativistic dilaton}
\end{align}
contains the dilaton $X$, the 2d Ricci scalar $R$, and arbitrary potential $U(X)$ and $V(X)$ that characterize the particular model.

A detailed discussion of the large-$c$ expansion of Einstein gravity within the pre-Non-Relativistic (PNR) setup \cite{Bergshoeff:2017btm,Hansen:2020pqs} is reviewed in Appendix \ref{app:C}. In what follows, we implement this expansion for the relativistic dilaton action \eqref{relativistic dilaton}.

As derived in Eq.~\eqref{c expansion of Ricci}, for a general 2d Lorentzian manifold, the Ricci scalar can be expanded as
\begin{align}
   R= \frac{1}{c^2}\,\big(-V^\m V^\n\ggtR{_\m_\n}|_{\gamma=0}-\ggtna{_\m}(V^\m \mathcal{K})\big)-2\big(\ggtna{_\m}A^\m+A^\m A_\m\big)\,.
\end{align}
Here, $\ggtR{_\m_\n}|_{\gamma=0}$ denotes the Ricci tensor associated with the minimal pre-Galilean connection, which is obtained by fixing the ambiguity parameter to $\gamma=0$ and is defined in Eqs.~\eqref{coe gali} and \eqref{Ricci def}. The symbol $\ggtna{_\m}$ represents the corresponding minimal covariant derivative, free of spatial torsion, constructed from the connection in Eq.~\eqref{coe gali} evaluated at $\gamma=0$. The quantities $V^\m, T_\m, \Theta^\m, \mathcal{E}_\m$ are components of the local Lorentzian vielbein and serve as the PNR variables introduced by local metric decomposition \eqref{pre g a} and \eqref{pre g b}. Furthermore, $\mathcal{K}=\mathcal{K}_{\m\n}\Theta^\m \Theta^\n$, with $\mathcal{K}_{\m\n}=-\mathcal{E}_{(\m}\mathcal{L}_{V}\mathcal{E}_{\n)}$, defines the pre-extrinsic curvature scalar, while $A_{\m}=\mathcal{L}_{V}T_{\m}=V^\n(\dd T)_{\n\m}$ specifies the pre-acceleration vector.

We then implement a large-$c$ expansion of the PNR variables
\begin{align}
    V^\m&=v^\m+\mathcal{O}(c^{-2}) & T_\m&=\tau_{\m}+\mathcal{O}(c^{-2}) &
    \Theta^\m &=\theta^\m+\mathcal{O}(c^{-2}) & \mathcal{E}_{\m}&=e_\m+\mathcal{O}(c^{-2})\,.
\end{align}
These field expansions induce the following large-$c$ expansion for these geometric data:
\begin{subeqs}
    A_{\m}&=a_\m +\mathcal{O}(c^{-2})& \quad\mathcal{K}&=K+\mathcal{O}(c^{-2})& \\
    \gtna{}&=\hat{\nabla}+\mathcal{O}(c^{-2})&\quad \ggtR{_\m_\n}&=\hat{R}_{\m\n}+\mathcal{O}(c^{-2})& \sqrt{-g}=c\det(\tau,e)+\mathcal{O}(c^{-1})\,.
\end{subeqs}
Inserting these large-$c$ expansions into the relativistic action \eqref{relativistic dilaton}, we obtain a Laurent series in powers of $c^{-2}$,
\begin{align}
I_{\mathrm{dila}}&=c^4I_{\mathrm{LO}}+c^2I_{\mathrm{NLO}}+\mathcal{O}(c^{0})
\end{align}
where leading and subleading contributions are given by
\begin{subeqs}
    I_{\mathrm{LO}}&=\frac{k_{\mathrm{rel}}}{4\pi}\int\dd^2x\det(\tau,e)\left[2X(\hat{\nabla}_\m a^\m+a^\m a_\m)+2V(X)-(\theta^\m \p_\m X)^2 U(X)\right]\\
    I_{\mathrm{NLO}}&=\frac{k_{\mathrm{rel}}}{4\pi}\int\dd^2x\det(\tau,e)\left[- Xv^\m v^\n \hat{R}_{\m\n}|_{\gamma=0}-X\hat{\nabla}_\m(v^\m K)+(v^\m \p_\m X)^2 U(X)\right]\,. \label{eq:NLOaction}
\end{subeqs}
Using the divergence identity \eqref{g div}, the leading action
can be written, up to a boundary term, as 
\begin{align}
	I_{\mathrm{LO}}=\frac{k_{\mathrm{rel}}}{4\pi}\int\dd^2x\det(\tau,e)
	\big[2(\theta^\mu a_\mu)(\theta^\nu\p_\nu X)+2V(X)-U(X)(\theta^\mu\p_\mu X)^2\big]\,.
\end{align}
We then introduce two auxiliary fields $\tilde\rho$ and $\tilde X_H$, finite and non-vanishing in the large-$c$ limit, and scale the potential according to
\begin{align}
V(X)=c^{-2}\tilde V(X)\,.
\label{eq:2dVscaling}
\end{align}
An equivalent auxiliary-field form of
the leading contribution is
\begin{eqs}
    I_{\mathrm{LO}}=\frac{k_{\mathrm{rel}}}{4\pi c^2}\int\dd^2x\,\det(\tau,e)
	\big[
	2\tilde V+\tilde\rho(\theta^\mu\p_\mu X)-\tilde X_H(\theta^\mu a_\mu)+c^{-2}\big(\frac12\tilde\rho\tilde X_H-\frac14U\tilde X_H^2\big)
	\big]\,. \label{eq:LOaction}
\end{eqs}
Eliminating the auxiliary fields gives
\begin{align}
	\tilde X_H=-2c^2\theta^\mu\p_\mu X\qquad \qquad
	\tilde\rho=2c^2\left[\theta^\mu a_\mu-U(X)\theta^\mu\p_\mu X\right] \label{eq:auxi}
\end{align}
We define the finite Galilean coupling to ensure a finite resulting action by
\begin{align}
    k=\lim_{c\to\infty}c^2k_{\mathrm{rel}}\,.
    \label{eq:kGal}
\end{align}
Combining the auxiliary-field expression of leading action \eqref{eq:LOaction} with the next-leading action \eqref{eq:NLOaction} and finally taking this limit, we obtain
\begin{align}
    I_{\mathrm{dil}}[X,\,\tilde{\rho},\,\tilde{X}_{H},\,\tau,\,e]&=\frac{k}{4\pi}\int\dd^2x\det(\tau,e)\big[-X\big( v^\m v^\n \hat{R}_{\m\n}|_{\gamma=0}+\hat{\nabla}_\m(v^\m K)\big)\nonumber\\
    &\quad\quad\quad\quad +\tilde{\rho} (\theta^\m a_\m)-\tilde{X}_{H}(\theta^\m \p_\m X)+2\tilde{V}+(v^\m \p_\m X)^2 U(X)\big]\,.
\end{align}
Using the 2d identities Eqs.~\eqref{2Ricci}, \eqref{omegahat}, \eqref{g div}, we verify that in 2d
\begin{align}
    v^\m v^\n \hat{R}_{\m\n}= v^\m v^\n R[\hat{\omega}]_{\m\n}=\hat{\nabla}_\m(v^\m K)\,.\label{electric limit}
\end{align}
Using Eq.~\eqref{electric limit}, the action above reproduces the second-order formulation of the 2d Galilean dilaton model \eqref{2nd action}, up to the redefinition of multipliers $\tilde{X}_H\rightarrow 2\XH$ and $\tilde{\r}\rightarrow 2\r$ only.

\subsubsection[Galilean limit of \texorpdfstring{$D$}{D}-dimensional Einstein gravity]{Galilean limit of \texorpdfstring{$\boldsymbol{D}$}{D}-dimensional Einstein gravity}
We next reverse the order of the two operations and take the large-$c$ limit directly in $D$-dimensional Einstein gravity before spherically reducing. The argument below applies to $D>3$.

At finite $c$, we write the metric in a generic ADM form,
\begin{align}
\dd s_D^2=-c^2N^2\dd t^2+H_{ij}(\dd x^i+N^i\dd t)(\dd x^j+N^j\dd t)\qquad i,j=1,\ldots,D-1\,.
\label{eq:DADM}
\end{align}
The $D$-dimensional Einstein--Hilbert action is
\begin{align}
I_{\rm EH}^{(D)}=\frac{c^3}{16\pi G_D}\int\dd^Dx\,\sqrt{-g}\,R\,.
\end{align}
In the variables of Eq.~\eqref{eq:DADM}, its bulk part takes the standard ADM form,
\begin{align}
I_{\rm EH}^{(D)}=\frac{c^4}{16\pi G_D}\int\dd t\,\dd^{D-1}x\,N\sqrt H\left[R[H]+\frac{1}{c^2}\left(\mathcal K_{ij}\mathcal K^{ij}-\mathcal K^2\right)\right]
\label{eq:DADMaction}
\end{align}
and
\begin{align}
\mathcal K_{ij}=\frac{1}{2N}\left(\p_tH_{ij}-\mathcal L_{\vec N}H_{ij}\right)\qquad \mathcal K=H^{ij}\mathcal K_{ij}\,.
\end{align}

We next restrict to spherically symmetric spacetimes,
\begin{align}
\dd s_D^2=-c^2N^2\dd t^2+E^2(\dd r+N^r\dd t)^2+\mathcal R^2(t,r)\dd\Omega_{D-2}^2
\label{eq:DwarpADM}
\end{align}
where $N$, $E$, $N^r$, and $\mathcal R$ depend only on $(t,r)$. We then introduce
\begin{align}
\tau=N\,\dd t\quad\qquad e=E(\dd r+N^r\dd t)\quad\qquad v=-\frac{1}{N}(\p_t-N^r\p_r)\quad\qquad \theta=\frac{1}{E}\p_r\,.
\end{align}
The contractions appearing in the action are
\begin{align}
\det(\tau,e)=NE\qquad\qquad \theta^\mu a_\mu=\frac{\p_rN}{NE}\qquad\qquad K=\frac{\p_tE-\p_r(EN^r)}{NE}\,.
\end{align}
For the metric defined in Eq.~\eqref{eq:DwarpADM}, the curvature and kinetic terms in Eq.~\eqref{eq:DADMaction} reduce to
\begin{align}
&R[H]
=\frac{(D-2)(D-3)}{\mathcal R^2}
-\frac{2(D-2)}{\mathcal R}\theta^\mu\p_\mu\left(\theta^\nu\p_\nu\mathcal R\right)
-\frac{(D-2)(D-3)}{\mathcal R^2}\left(\theta^\mu\p_\mu\mathcal R\right)^2
\\
&\mathcal K_{ij}\mathcal K^{ij}-\mathcal K^2
=\frac{2(D-2)K}{\mathcal R}v^\mu\p_\mu\mathcal R
-\frac{(D-2)(D-3)}{\mathcal R^2}\left(v^\mu\p_\mu\mathcal R\right)^2\,.
\end{align}
After integrating over the sphere, the bulk action becomes
\begin{align}
\begin{aligned}
I_{\rm EH}^{(D)}=\frac{c^4\Omega_{D-2}}{16\pi G_D}\int&\dd^2x\,\det(\tau,e)\,\mathcal R^{D-2}\Bigg[
\frac{(D-2)(D-3)}{\mathcal R^2}
+2(D-2)(\theta^\mu a_\mu)\frac{\theta^\nu\p_\nu\mathcal R}{\mathcal R}\\
&+(D-2)(D-3)\left(\frac{\theta^\mu\p_\mu\mathcal R}{\mathcal R}\right)^2\\
&+\frac{1}{c^2}\left(2(D-2)K\frac{v^\mu\p_\mu\mathcal R}{\mathcal R}-(D-2)(D-3)\left(\frac{v^\mu\p_\mu\mathcal R}{\mathcal R}\right)^2\right)\Bigg]\,.
\end{aligned}
\label{eq:DwarpAction}
\end{align}

To obtain a non-trivial Galilean limit, we require both the spherical potential and the time-dependent terms in Eq.~\eqref{eq:DwarpAction} to remain finite and non-zero. We write $\mathcal R=c^z\bar{\mathcal R}$ and $G_D=c^\eta\bar G_D$. The two parts then scale as $c^{4+(D-4)z-\eta}$ and $c^{2+(D-2)z-\eta}$, respectively. For $D>3$, their simultaneous finiteness fixes
\begin{align}
\mathcal R=c\,\bar{\mathcal R}\qquad\qquad G_D=c^D\,\bar G_D
\label{eq:DwarpScaling}
\end{align}
where the barred variables remain finite at large $c$. The remaining spatial-gradient terms are of order $c^2$ and require the same auxiliary-fields insertion encountered in the direct 2d limit as shown in Eq.~\eqref{eq:auxi}.

 At finite $c$, the standard spherical reduction defines
\begin{align}
X=\left(\frac{\lambda_{\rm rel}\mathcal R}{D-2}\right)^{D-2} \qquad\qquad
k_{\rm rel}=\frac{\Omega_{D-2}}{4G_D}\left(\frac{D-2}{\lambda_{\rm rel}}\right)^{D-2}\,. \label{eq:rescaleXlambda}
\end{align}
Keeping $X$ finite under Eq.~\eqref{eq:DwarpScaling} requires
\begin{align}
\lambda_{\rm rel}=\frac{\lambda}{c}
\end{align}
with $\lambda$ fixed on the large-$c$ limit. Together with $G_D=c^D\bar G_D$, this gives
\begin{align}
k_{\rm rel}=\frac{1}{c^2}\frac{\Omega_{D-2}}{4\bar G_D}\left(\frac{D-2}{\lambda}\right)^{D-2}\,.
\end{align}
Thus the finite Galilean coupling is
\begin{align}
k\equiv\lim_{c\to\infty}c^2k_{\rm rel}=\frac{\Omega_{D-2}}{4\bar G_D}\left(\frac{D-2}{\lambda}\right)^{D-2}\,.
\end{align}
The distinction between $k_{\rm rel}$ and $k$ is the same as in the direct 2d limit of Eq.~\eqref{eq:kGal}.

The spherical-reduction potentials at finite $c$ are
\begin{align}
U_D(X)&=-\frac{D-3}{(D-2)X}\qquad\qquad
V_D^{\rm rel}(X)=\frac{\lambda_{\rm rel}^2(D-3)}{2(D-2)}X^{\frac{D-4}{D-2}}\,.
\end{align}
Using $\lambda_{\rm rel}=\lambda/c$, the second relation becomes
\begin{align}
V_D^{\rm rel}(X)=\frac{1}{c^2}V_D(X)\qquad\qquad
V_D(X)=\frac{\lambda^2(D-3)}{2(D-2)}X^{\frac{D-4}{D-2}}\,.
\end{align}
This is precisely the potential scaling \eqref{eq:2dVscaling} required in the direct 2d limit, with $\tilde V(X)=V_D(X)$. Here, however, it is not imposed independently: it follows from the higher-dimensional scaling $\mathcal R=c\bar{\mathcal R}$.

Expressed in terms of the finite variables $X$, $k$, and $V_D$, Eq.~\eqref{eq:DwarpAction} becomes
\begin{align}
\begin{aligned}
I_{\rm EH}^{(D)}=\frac{k}{4\pi}\int\dd^2x\,\det(\tau,e)\Bigg[&
c^2\left[2(\theta^\mu a_\mu)(\theta^\nu\p_\nu X)-U_D(X)(\theta^\mu\p_\mu X)^2\right]\\
&+2V_D(X)+2K(v^\mu\p_\mu X)+U_D(X)(v^\mu\p_\mu X)^2\Bigg]\,
\end{aligned}
\label{eq:DwarpLargeC}
\end{align}
up to a boundary term.
By introducing the auxiliary fields $\tilde\rho$ and $\tilde X_H$, we write the corresponding $D$-dimensional EH action as
\begin{align}
\begin{aligned}
I_{\rm EH}^{(D)}=\frac{k}{4\pi}&\int\dd^2x\,\det(\tau,e)\Bigg[
2V_D(X)+2K(v^\mu\p_\mu X)+U_D(X)(v^\mu\p_\mu X)^2\\
&+\tilde\rho\,\theta^\mu\p_\mu X-\tilde X_H\,\theta^\mu a_\mu+\frac{1}{c^2}\left(\frac12\tilde\rho\,\tilde X_H-\frac14U_D(X)\tilde X_H^2\right)\Bigg]\,.
\end{aligned}
\label{eq:DwarpAux}
\end{align}
The auxiliary equations are solved as
\begin{align}
\tilde X_H=-2c^2\theta^\mu\p_\mu X\qquad \tilde\rho=2c^2\left[\theta^\mu a_\mu-U_D(X)\theta^\mu\p_\mu X\right]
\end{align}
and eliminating the auxiliary fields reproduces Eq.~\eqref{eq:DwarpLargeC}.

Keeping $\tilde\rho$ and $\tilde X_H$ finite as $c\to\infty$, the $c^{-2}$ terms in Eq.~\eqref{eq:DwarpAux} drop out and the auxiliary fields become Lagrange multipliers for the spatial constraints. Defining $\tilde\rho=2\rho, \tilde X_H=2\XH$
we obtain
\begin{align}
\begin{aligned}
I_{\rm G}^{(D)}\big|_{\rm sph}=\frac{k}{2\pi}\int\dd^2x\,\det(\tau,e)\Bigg[
&V_D(X)+K(v^\mu\p_\mu X)+\frac12U_D(X)(v^\mu\p_\mu X)^2\\
&+\rho\,\theta^\mu\p_\mu X-\XH\,\theta^\mu a_\mu\Bigg]\,.
\end{aligned}
\label{eq:DGsph}
\end{align}
Using 
\begin{align}
K(v^\mu\p_\mu X)=-X\hat\nabla_\mu(v^\mu K)+\hat\nabla_{\mu}(XK v^\mu)\,,\qquad v^\mu v^\nu\hat R_{\mu\nu}=\hat\nabla_\mu(v^\mu K)\,,
\end{align}
Eq.~\eqref{eq:DGsph} reproduces the second-order Galilean dilaton Lagrangian \eqref{2nd action} with $U(X)=U_D(X)$ and $V(X)=V_D(X)$.

Spherical reduction and the Galilean limit therefore give the same 2d Galilean theory. In the direct 2d  derivation, the scalings $V\to c^{-2}\tilde V$ and $k_{\rm rel}\sim c^{-2}$ were imposed explicitly. In the higher-dimensional derivation, both follow from the scaling of $\mathcal R$ and $G_D$ through the spherical-reduction dictionary. Thus such two routes with the opposite orders of the two operations in Fig.~\ref{derivation_path} commute.

\subsection{Constant dilaton vacua}\label{sec:2.4}

A constant dilaton configuration, $X=X_0=\textrm{const.}$, satisfies Eq.~\eqref{eom:dilaton} only if $\XP=0$. 
The auxiliary equation~\eqref{eom:auxiliary} then imposes the algebraic condition
\begin{equation}
\mathcal V(X_0,0)=V(X_0)=0\,.
\end{equation}
Constant dilaton vacua therefore exist only at zeros of the dilaton potential and need not exist in a given model. For example, a non-zero constant potential $V(X)=\Lambda_{\rm CGHS}$, as in the Galilei--CGHS model, admits no such vacuum. Equation~\eqref{eom:casimir} further reduces to $\dd\XH=0$, leaving $\XH$ as an arbitrary constant.

At $X=X_0$ and $\XP=0$, the remaining field equations become
\eq{
R[G]=-\Lambda\,\tau\wedge e\qquad\qquad R[H]=0\qquad\qquad R[P]=0
}{eq:vacua}
where
\begin{align}
\Lambda:=\left.\p_X\mathcal V(X,\XP)\right|_{X=X_0,\XP=0}=V'(X_0)\,.
\end{align}
Thus the constant dilaton model has constant Galilean curvature and vanishing intrinsic and spatial torsion. The latter follows from $\p_P\mathcal V=U(X)\XP$ for the class of potentials \eqref{def mathcal V}.

Since $R[H]=\dd\tau=0$, we may locally choose
\begin{align}
\tau=\dd t\,.
\end{align}
In a gauge with $e_t=\omega_t=0$, write $e=e_r(t,r)\,\dd r$ and $\omega=\omega_r(t,r)\,\dd r$. The equations $R[P]=0$ and $R[G]=-\Lambda\tau\wedge e$ then give
\begin{align}
\omega_r=\p_t e_r\qquad\qquad \p_t^2e_r+\Lambda e_r=0\,.
\end{align}
The local solutions are therefore
\begin{alignat}{3}
&\text{Flat}\quad \Lambda=0:\qquad
&&e=(A(r)t+B(r))\,\dd r
\\
&\text{AdS}\quad \Lambda>0:\qquad
&&e=\left[A(r)\sin(\sqrt{\Lambda}t)+B(r)\cos(\sqrt{\Lambda}t)\right]\dd r
\\
&\text{dS}\quad \Lambda<0:\qquad
&&e=\left[A(r)\sinh(\sqrt{-\Lambda}t)+B(r)\cosh(\sqrt{-\Lambda}t)\right]\dd r\,.
\end{alignat}
Here $A(r)$ and $B(r)$ are arbitrary integration functions. The dependence on $t$ is fixed entirely by the sign of $\Lambda$.\footnote{We use the sign of the Galilean Ricci scalar to distinguish the two branches: $R<0$ is referred to as AdS and $R>0$ as dS. This convention is compatible with the corresponding Lorentzian JT theories under the Galilean limit. In the convention of \eqref{relativistic dilaton}, Lorentzian AdS$_2$ JT gravity has $V(X)=+X/\ell^2$, while dS$_2$ JT gravity has $V(X)=-X/\ell^2$ \cite{Grumiller:2020elf}. Accordingly, the Galilean theory gives $R=-2/\ell^2$ and $R=+2/\ell^2$, respectively; see also the GJT potential \eqref{GJT potential} in Section~\ref{sec:GJTC}.}

The constant dilaton vacua form a separate algebraic branch of the solution space. We now turn to the generic linear-dilaton sector.

\subsection{Linear dilaton vacua}
\label{sec:2.5}
In this Subsection, we derive the exact solutions of the general Galilean dilaton gravity model~\eqref{2d first order action} with quartic potential~\eqref{def mathcal V}.
We start with the simple case with vanishing $U(X)$ and then extend the construction to a general model, by using the previously established Weyl-rescaling.

The intrinsic torsion constraint \eqref{eom:intrinsic_torsion}, $\dd \tau = 0$, allows us to choose locally
\begin{align}
    \tau = \dd t\,.
\end{align}
In this gauge, the scalar field equations \eqref{eom:dilaton} and \eqref{eom:auxiliary} lead to time-evolution equations,
\begin{align}
    \dd X = -\XP\, \dd t\qquad\qquad \dd \XP = V(X)\, \dd t\,.
\end{align}
 These imply that $X$ and $\XP$ are functions of time $t$ only. For $U=0$, the Casimir invariant $M$ relates these two fields algebraically by integrating these two equations
\begin{align}
    \frac{1}{2}\,\XP^2 + w(X) = M \qquad \text{with} \qquad w(X) \equiv \int^X V(y)\dd y\, .
\end{align}
This determines the trajectory of the auxiliary field as
\eq{\XP(X) = \pm \sqrt{2M - 2w(X)}}{}
And the dilaton is obtained by integration 
\eq{
  \pm \frac{\dd X}{\sqrt{2M-2w(X)}}=\dd t\,.
}{solution X}
Here, we have assumed we are working on the patches where $\XP\ne0$; below, we will relax this assumption and show that the singular loci $\XP=0$ play the central role in our definition of Galilean manifold.

To determine the remaining gauge fields, we employ Galilean boosts to fix $\XH = 0$. Then the Casimir equation \eqref{eom:casimir} becomes
\begin{align}
  \omega = \frac{V(X)}{\XP}\, e\,.
\end{align}
Substituting this result into the spatial torsion equation \eqref{eom:spatial_torsion}, $\dd e + \omega \wedge \tau = 0$, yields
\begin{align}
    \dd e - \frac{\dd \ln \XP}{\dd t}\, \dd t \wedge e = 0\,.
\end{align}
The general solution to this equation is $e = \XP(t)\, \dd r$.

In summary, the complete solution in the gauge $\XH=0$ for the $U=0$ model is given by
\begin{subeqs}
   \XH &= 0 \quad (\text{by gauge}) & \tau &= \dd t \\
        \XP &= \pm\sqrt{2M-2w(X)} & e &= \XP\, \dd r \\
        X&\;\text{given by integration~\eqref{solution X}} & \omega &= V(X)\, \dd r\,.
\end{subeqs}
The explicit expressions for metric and curvature 2-form are 
\begin{align}
    h&=\XP^2\,\dd r \otimes \dd r &
    R[G]&=- \frac{\dd V}{\dd X}\, \XP\, \dd t\wedge \dd r =\frac{\dd V}{\dd X}\,\dd X\wedge\dd r\,.
\end{align}

We can extend the $U=0$ solution above to theories with arbitrary kinetic coupling $U(X)$ using Weyl rescalings with the Weyl factor $Q(X)$ defined in Eq.~\eqref{def Q}. The Weyl-transformed Casimir equation
\eq{
	{
		\frac{1}{2}e^{Q(X)}\XP^2+\int^X e^{Q(y)}V(y)\,\dd y=M
	}
}{Casimir eq}
determines the auxiliary field
\begin{equation}
    \XP=\pm  \sqrt{ 2e^{-Q(X)}\left( M - w(X) \right)}
\end{equation}
where $w$ is defined by 
\begin{equation}
    w(X)=\int^{X} e^{Q(y)} V(y)\, \dd y
\end{equation}
and the dilaton
\eq{
   \pm \frac{\dd X}{\sqrt{2e^{-Q(X)}\left(M - w(X)\right)}}=\dd t\,.
}{eq:dilaton}
Here and in the following, we assume that $e^{Q(X)}$ is finite and non-vanishing on the patch under consideration; loci where the Weyl factor vanishes or diverges will be discussed separately. The exact solution is given by
\begin{align}
	&\tau=\dd t\qquad\qquad
	e=e^{Q(X)}\XP\,\dd r\qquad\qquad
	\omega=e^{Q(X)}\mathcal{V}\,\dd r\,.\label{tau e solution}\\
	&h_{\m\n}\,\dd x^\m\dd x^\n
	=e^{2Q(X)}\XP^2\,\dd r\otimes\dd r
	\qquad\qquad
	h^{\m\n}\p_\m\p_\n
	=e^{-2Q(X)}\XP^{-2}\,\p_r\otimes\p_r\,.
	\label{cometric solution}
\end{align}
From this expression, the cometric becomes singular at points where the auxiliary field vanishes, $\XP(t)=0$. The physical nature and geometric implications of this singularity will be investigated in detail in Section \ref{sec:3}.

Furthermore, the spacetime is characterized by a vanishing intrinsic torsion, $R[H] = \dd \tau = 0$, which ensures the existence of the temporal foliation of spacetime. However, the Galilean curvature 2-form, is generally non-zero and is determined by the dilaton potential,
\begin{align}
    R[G] = -e^{Q(X)}\, \XP\, \p_X\mathcal{V}\,\dd t \wedge \dd r=e^{Q(X)}\,\p_X\mathcal{V}\,\dd X\wedge\dd r
\end{align}
and the corresponding scalar curvature is defined as 
\begin{equation}
    R\equiv-2v^\m v^\n R_{\m\n}= -2 v^\m \theta^\n R[G]_{\m\n}=-2\p_X \mathcal{V}
\end{equation}
where $R_{\m\n}$ is the Ricci tensor on the $2d$ spacetime defined by Eq.~\eqref{Ricci def}.

\subsection{Galilean Birkhoff theorem}
\label{sec:2.6}

We now determine the local Killing algebra of the constant dilaton vacua and general non-constant solution respectively. Following the definition in Eq.~\eqref{killing b}, additionally imposing the invariance of the dilaton, we require
\begin{align}
\mathcal L_\xi\tau=0\qquad\qquad \mathcal L_\xi h^{\mu\nu}=0\qquad\qquad \mathcal L_\xi X=0\,.
\end{align}
In two dimensions, the second condition is equivalent to $\mathcal L_\xi\theta=0$.

For a constant dilaton vacuum, $\mathcal L_\xi X=0$ is automatic. Equation~\eqref{eq:vacua} gives $\dd\tau=0$, so locally $\tau=\dd t$. Writing $e=e_r(t,r)\,\dd r$, Eq.~\eqref{eq:def theta} gives $\theta=e_r^{-1}\p_r$. Introducing
\begin{align}
x(t,r)=\int^r e_r(t,r')\,\dd r'
\end{align}
the Killing vector is given by
\begin{align}
\xi=\epsilon\,\p_t+f(t)\p_x
\end{align}
where $\epsilon$ is constant and $f(t)$ arbitrary. In the original $(t,r)$ coordinates, we have
\begin{equation}
\xi_H=\p_t-\frac{\p_t x(t,r)}{e_r(t,r)}\,\p_r\qquad\qquad \xi_f=\frac{f(t)}{e_r(t,r)}\,\p_r\,.\label{eq:consKV}
\end{equation}
Their brackets
\begin{align}
[\xi_H,\xi_f]=\xi_{\dot f}\qquad\qquad [\xi_{f_1},\xi_{f_2}]=0
\end{align}
yield the infinite-dimensional Killing algebra
\begin{align}
\mathfrak g_{\rm CDV}=\mathbb R\ltimes C^\infty(\mathbb R)
\end{align}
for the constant-dilaton solutions, with $C^\infty(\mathbb R)$ an Abelian ideal. The generators $\xi_H$, $\xi_{f=1}$, and $\xi_{f=t}$ give the $\mathfrak{gal}(2)$ subalgebra.

For the generic linear-dilaton solutions of Section~\ref{sec:2.5}, we restrict to patches with $\XP\neq0$. Equations~\eqref{eom:dilaton} and \eqref{tau e solution} give $\dot X=-\XP$, so $\mathcal L_\xi X=0$ fixes $\xi^t=0$. Moreover, Eqs.~\eqref{tau e solution} and \eqref{eq:def theta} give $\theta=e^{-Q(X)}\XP^{-1}\p_r$, whose coefficient depends only on $t$. Hence $\mathcal L_\xi\theta=0$ implies $\p_r\xi^r=0$, 
\eq{
\xi_f=f(t)\p_r
}{supertranslation}
with arbitrary $f(t)$. These are precisely the time-dependent spatial translations appearing in the general Galilean Killing vectors \eqref{flat killing}. The non-constant dilaton excludes the additional generator $\xi_H$ present in~Eq.\eqref{eq:consKV} in the constant-dilaton vacua. These transformations are the Galilean analogue of BMS supertranslations, with the arbitrary function depending on time instead of angular coordinates, and the translation is in space instead of in time.

Thus a general Galilean dilaton theory admits a spatial Killing vector, while the constant-dilaton vacua have an enhanced Killing algebra. This is the local Galilean analogue of the Birkhoff theorem.

\section{G-tachy manifolds and G-tachydynamics }\label{sec:3}

In this Section, we introduce 2d G-tachy manifolds and define the associated \textit{Galilean tachydynamics} at the extremal surface $X_{\textrm{\tiny P}}=0$. We relax $\XP\neq0$ made in Section~\ref{sec:2.5} and consider isolated $\XP=0$, where the Galilean structure degenerates. We study the geometry of these solutions with compact spatial Killing orbits $r\sim r+\beta$, identify Galilean extremal surfaces, and derive the associated first law from covariant phase-space charges.

The terminology G-tachy reflects the fact that in the general Galilean dilaton theory, the Killing vector guaranteed by the Galilean Birkhoff theorem generates spatial rather than temporal translations. Its conserved charge is therefore a momentum $P$ rather than an energy $E$, and the quantity conjugate to entropy $S$ is the \textit{Galilei velocity} $\vmakro$ instead of the temperature. Using the covariant phase-space formalism, we derive the corresponding entropy and the first law of Galilean tachydynamics, $\delta P=\vmakro\,\delta S$.

\subsection{G-tachy manifolds}

The generic linear-dilaton solutions in Section~\ref{sec:2.5} assume $\XP\neq0$. We now relax it. Let $\mathcal M$ be a compact, connected, smooth and orientable two-manifold, and define its regular Galilean region by
\begin{align}
\mathcal M_{\rm reg}\equiv\{p\in\mathcal M\,|\,\XP(p)\neq0\}\,.
\end{align}

By the Galilean Birkhoff theorem of Section~\ref{sec:2.6}, the generic linear-dilaton solutions admit a spatial Killing vector $\xi=\p_r$. In direct analogy with the compact orbit used in the Carrollian thermal construction~\cite{Ecker:2023uwm}, we set
\begin{align}
r\sim r+\beta\qquad \beta>0\,.
\end{align}

From Eq.~\eqref{tau e solution}, $e=e^{Q(X)}\XP\,\dd r$, while Eq.~\eqref{cometric solution} gives $h^{rr}=e^{-2Q}\XP^{-2}$. For finite and non-vanishing $e^Q$, $\theta^\mu$ and hence the $h^{rr}$ blow up as $\XP\to0$. This degeneration cannot be removed by a Galilean boost. If the underlying manifold is smooth at an isolated $X_{\textrm{\tiny P}}=0$, a neighborhood of that point has disk topology.

For a closed orbit $\gamma$ generated by $\xi=\p_r$ surrounding an isolated $X_{\textrm{\tiny P}}=0$, the boost holonomy is
\begin{align}
H_\gamma=\exp\left[-\oint_\gamma\omega\,G\right]=\exp\left[-\beta e^{Q(X)}\mathcal V(X,\XP)\,G\right]
\end{align}
where $G$ is the Galilean boost generator. At a $X=X_i$ with $\XP(X_i)=0$, $e^{Q(X_i)}\mathcal{V}(X_i,0)=w^\prime(X_i)$. Following the Carrollian prescription of Ref.~\cite{Ecker:2023uwm}, we impose the regularity condition at each $X_P(X_i)=0$, which fixes the corresponding positive $\beta_i$ to
\begin{align}
 \beta_i=\frac{2\pi}{|w'(X_i)|}
\label{eq:capholo}
\end{align}
This condition removes an additional conical-type defect but does not regularize the intrinsic Galilean degeneration. A simple root, $w'(X_i)\neq0$, which leads to a finite $\beta$, therefore admits a finite cap period.

We define the Galilei velocity and its signed counterpart by
\begin{align}
\vmakro_i=\beta_i^{-1}=\frac{|w'(X_i)|}{2\pi}\qquad\qquad \tilde{\vmakro}_i=\frac{w'(X_i)}{2\pi}\,.
\end{align}
By writing $\ell_\xi=e^Q\XP$, we define the Galilean surface gravity by
\begin{align}
\left.\nabla_\mu\ell_\xi\right|_{X_i}=\tilde{\kappa}_i\,\tau_\mu\,.
\label{eq:surfacegravity}
\end{align}
Using Eqs.~\eqref{eom:dilaton} and \eqref{eom:auxiliary} together with $Q'=U$, one finds
\begin{align}
\tilde{\kappa}_i=e^{Q(X_i)}V(X_i)=w'(X_i)\qquad\qquad \tilde{\vmakro}_i=\frac{\tilde{\kappa}_i}{2\pi}\,.
\end{align}

If $\XP$ is nowhere vanishing, the compact spatial orbits do not collapse. The possible topologies are a cylinder and a torus. The torus is incompatible with the generic non-constant linear-dilaton equation of motion~\eqref{eom:dilaton}. Hence the non-degenerate topology with $\XP\neq0$ is cylindrical.

The Casimir relation gives
\begin{align}
\frac12e^{Q(X)}\XP^2=M-w(X)\geq0\,.
\end{align}
A simple root $w(X_i)=M$ changes the sign of $M-w(X)$ and therefore lies at an endpoint of a connected physical interval where $X_{\textrm{\tiny P}}^2\ge 0$. Thus one such endpoint together with a non-collapsing boundary gives a disk, while two such endpoints give an $S^2$.

At a multiple root $X_*$ of finite order, $\omega^\prime(X_*)=0$. Equation~\eqref{eq:dilaton} then gives
\begin{align}
\Delta t=\pm\int\frac{e^{Q(X)/2}}{\sqrt{2[M-w(X)]}}\,\dd X
\end{align}
which diverges as $X\to X_*$ when $e^{Q(X_*)}$ is finite and non-vanishing. The multiple root is therefore reached only at infinite Galilean time and defines an asymptotic boundary. Since $w'(X_*)=0$, no finite $\beta$ satisfies Eq.~\eqref{eq:capholo}. 

For a connected interval bounded by two simple roots $X_-<X_+$, with $M-w(X)>0$ between them,
\begin{align}
w'(X_-)<0\qquad\qquad w'(X_+)>0\,.
\end{align}
The two poles therefore carry opposite signed surface gravities, as indicated in Fig.~\ref{fig:phaseint}. The same  radial identification $r\sim r+\beta$ at two simple roots requires
\begin{align}
|w'(X_-)|=|w'(X_+)|\,.
\label{eq:s2match}
\end{align}
This is a global matching condition for models which admit an interval bounded by two simple roots of $X_{\textrm{\tiny P}}(X)=0$.

For such family of solutions labeled by continuous mass $M$, the implicit function theorem gives
\begin{align}
\frac{\dd}{\dd M}(X_-+X_+)=\frac{1}{w'(X_-)}+\frac{1}{w'(X_+)}=0\,.
\end{align}
Thus, $X_-+X_+=2X_*$ for some $M$-independent $X_*$. Since $w(X_-)=w(X_+)=M$, this implies
\begin{align}
X_*\pm x(M)=X_{\pm}(M)
\end{align}
which is the reflection symmetry of such class of solutions. If the two roots coincide as $M\to M_*$, then $X_\pm\to X_*$, $M_*=w(X_*)$, and $w'(X_*)=0$. Since $w(X_\pm)=M>M_*$, $X_*$ is a local minimum of $w$. Moreover, $w'(X_*)=e^{Q(X_*)}V(X_*)=0$ implies $V(X_*)=0$, so $X=X_*$ and $\XP=0$ define a constant-dilaton vacuum with  $M_*$. Thus, this family of solutions can be viewed as emerging from the constant-dilaton vacuum at $M=M_*$ with vanishing Galilean velocity $\vmakro_{*}$.

Conversely, models satisfying such matching condition \eqref{eq:s2match} can be constructed by choosing a regular $Q(X)$ and a reflection symmetric  $w(X)$ about $X_*$. If $w(X)=M$ has two symmetric simple roots with $M-w(X)>0$ between them, the matching condition follows automatically. A simple example is
\begin{align}
X_*&=0& Q(X)&=0& w(X)&=\frac{X^2}{2\ell^2}
\end{align}
which gives $U=0$ and $V=X/\ell^2$, namely the Galilei--Jackiw--Teitelboim model discussed in Section~\ref{sec:GJTC}.

Actually, such matching conditions need not hold for any two simple roots bounded a physical interval. Consider the Galilean Reissner--Nordstr\"om model obtained from the spherically reduced Einstein--Maxwell theory~\cite{Grumiller:2002nm} by the large-$c$ scaling discussed in Section~\ref{sec:2.3}. The corresponding potential is
\begin{align}
X=\rho^2\qquad\qquad w\bigl(X(\rho)\bigr)=\rho+\frac{q^2}{\rho}\qquad\qquad M=2\mu\,.
\end{align}
For $0<|q|<\mu$, the two simple roots and their corresponding $w^\prime$ are
\begin{align}
\rho_\pm&=\mu\pm\sqrt{\mu^2-q^2}&
|w'(X_\pm)|&=\frac{\sqrt{\mu^2-q^2}}{\rho_\pm^2}\,.
\end{align}
Since $\rho_+\neq\rho_-$, the two endpoints require different $\beta$ determined by Eq.~\eqref{eq:capholo}. Although both spatial orbits collapse, no  single global $\beta$ satisfies the regularity condition at both endpoints.

If $e^Q$ vanishes or diverges, $\XP=0$ alone no longer determines the orbit geometry, holonomy parameter, or curvature, which must be read from the full solution. In the Galilei--Schwarzschild--Tangherlini solution of Section~\ref{sec:GSTC}, for example, $e^Q$ diverges at $X=0$ together with the holonomy parameter and Ricci scalar, and the endpoint is a curvature singularity.

The resulting structures are summarized in Table~\ref{tab:Gstructures}.
\begin{table}[t]
\centering
\small
\begin{tabular}{c|cccc}
\hline
Roots of $M-w$ & None & One simple & Two simple & Multiple \\
\hline
Orbit behavior & No finite collapse & One collapsing end & Two collapsing ends & Asymptotic end \\
Cap period & Unfixed & $\beta_i$ & $\beta_-,\,\beta_+$ & No finite value \\
\hline
\end{tabular}
\caption{Spatial orbit behavior for finite and non-vanishing $e^Q$ at the roots. Two simple roots admit a common regular cap period only if $\beta_-=\beta_+$.}
\label{tab:Gstructures}
\end{table}

\subsection{Galilean extremal surface}

The locus $\XP=0$ also admits an algebraic condition in the PSM formulation. Under a pure Galilean boost, $\lambda_G=\lambda$ and $\lambda_H=\lambda_P=0$, Eqs.~\eqref{psm symm X}--\eqref{gauge e} give
\begin{align}
\delta_\lambda X&=0
&
\delta_\lambda\XP&=0
&
\delta_\lambda\XH&=\lambda\XP\,.
\end{align}
Thus the PSM target-space point is fixed under Galilean boosts precisely at
\begin{align}
\XP=0\,.
\end{align}
Using Eq.~\eqref{XP=vpX}, the same condition gives
\begin{align}
v^\mu\p_\mu X=0
\end{align}
so the dilaton is stationary along the Galilean time flow. We refer to this boost-fixed locus as the \textit{Galilean extremal surface}. The condition is independent of the form of the potential and identifies the same locus that appeared geometrically in the preceding Subsection. It is the Galilean analogue of the boost-invariant criterion used for Carroll extremal surfaces~\cite{Ecker:2023uwm}.

\subsection{G-tachydynamics}
\label{sec:tachydynamics}

By the Galilean Birkhoff theorem of Section~\ref{sec:2.6}, every generic linear-dilaton solution admits a spatial Killing vector. We take $\xi=\p_r$. The corresponding PSM charge variation is~\cite{Ecker:2023uwm}
\begin{align}
    \cancel{\delta}\mathcal Q_\lambda=\frac{k}{2\pi}\lambda_I\delta X^I\,.
    \label{eq:PSMcharge}
\end{align}
The on-shell identification $\lambda_I=\xi^\mu A_{I\mu}$ together with the Casimir relation \eqref{Casimir eq} gives
\begin{align}
\cancel{\delta}\mathcal Q_{\p_r}=\frac{k}{2\pi}\delta M\,.
\end{align}
We define the conserved spatial momentum
\begin{align}
P\equiv\mathcal Q_{\p_r}=\frac{k}{2\pi}M\,.
\label{eq:Gmomentum}
\end{align}
Unlike an energy, the momentum $P$ generated by the radial Killing vector can be negative. Thus, $M$ need not be positive in the general theory, although particular models may impose such an additional restriction.

We restrict in the following of this Subsection to regular Galilean extremal surfaces at simple roots $w(X_0)=M$, with finite and non-vanishing $e^{Q(X_0)}$ and $w'(X_0)\neq0$. The Galilean surface gravity defined in Eq.~\eqref{eq:surfacegravity} is then
\begin{align}
\tilde{\kappa}_0=w'(X_0)\,.
\end{align}

Following the Iyer–Wald formalism~\cite{Wald:1993nt,Iyer:1994ys}, we normalize the charge at the extremal surface by $|\tilde{\kappa}_0|$. For $\xi=\p_r$, the induced boost parameter is $\lambda_G=\tilde{\kappa}_0$ at $\XP=0$, while the remaining contribution to Eq.~\eqref{eq:PSMcharge} vanishes there. We therefore define
\begin{align}
\delta S\equiv\frac{2\pi}{|\tilde{\kappa}_0|}\left.\cancel{\delta}\mathcal Q_\xi\right|_{X_0}
=k\,\operatorname{sgn}(\tilde{\kappa}_0)\,\delta X_0\,.
\label{eq:Gentropyvar}
\end{align}
Here, the surface gravity $\tilde{\kappa}_0=w'(X_0)$ can be positive or negative. Choosing a reference position $X_{\rm ref}$ on the physical side of the root where $M-w(X)>0$, and choosing $S(X_{\rm ref})=0$, we obtain
\begin{align}
S=k\,\operatorname{sgn}(\tilde{\kappa}_0)(X_0-X_{\rm ref})
=k|X_0-X_{\rm ref}|
\label{eq:Gentropy}
\end{align}
which is non-negative for $k>0$. For a matched $S^2$ family where $w^\prime(X_+)=-w^\prime(X_-)$, the $M$-independent $X_*$ provides a natural reference position.

\begin{figure}[t]
\centering
\includegraphics[width=0.72\textwidth]{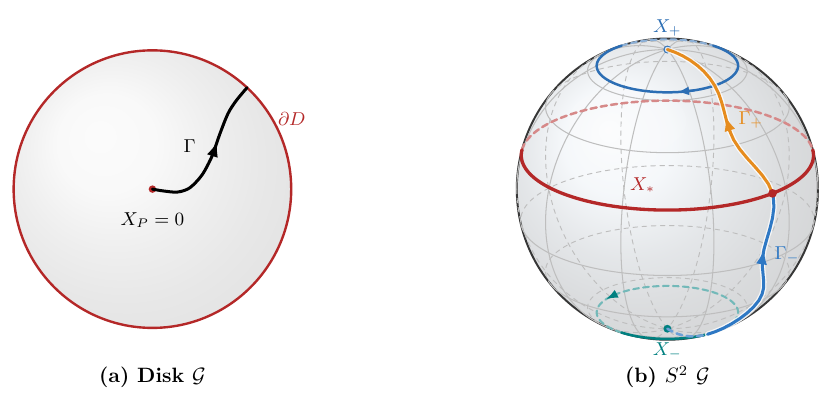}
\caption{Integration regions for the phase-space derivation. Left: a local disk bounded by a regular spatial orbit $\partial D$. Right: an $S^2$ solution divided at the regular orbit $X=X_*$ into two disk sectors with integration curves $\Gamma_\pm$. The two poles inherit opposite orientations from the two disk sectors, consistent with the opposite signs of the signed surface gravities, $\tilde{\kappa}_-<0$ and $\tilde{\kappa}_+>0$.}
\label{fig:phaseint}
\end{figure}

The first law follows from conservation of the exact-symmetry charge. For a local disk $D$, let $\partial D$ be any regular spatial orbit surrounding the extremal surface $X_{\textrm{\tiny P}}(X_0)=0$ and let $\Gamma$ connect $X_0$ to an arbitrary point on $\partial D$. For the case with two simple roots $X_\pm$ bounded a physical interval, we choose a regular orbit $X=X_{*}$ between them. The integral curves $\Gamma_\pm$ connect this orbit to the two endpoints, as shown in Fig.~\ref{fig:phaseint}.

Since $\xi=\p_r$ is an exact symmetry, the covariant phase-space identity gives $\dd\cancel{\delta}\mathcal Q_\xi=0$ on-shell~\cite{Ruzziconi:2020wrb}. Integration along $\Gamma$ therefore yields
\begin{align}
0=\int_\Gamma\dd\cancel{\delta}\mathcal Q_\xi
=\left.\cancel{\delta}\mathcal Q_\xi\right|_{\partial D}
-\left.\cancel{\delta}\mathcal Q_\xi\right|_{X_0}\,.
\end{align}
Using Eq.~\eqref{eq:Gentropyvar} and $\vmakro=|\tilde{\kappa}_0|/(2\pi)$, the integration above gives
\begin{align}
\delta P=\vmakro\,\delta S\,.
\label{eq:firstlaw}
\end{align}
For the case with two simple roots, the same argument separately applied to $\Gamma_\pm$ gives
\begin{align}
\delta P=\vmakro_\pm\,\delta S_\pm\,.
\end{align}
The two Galilei velocities $\vmakro_\pm=\beta_\pm^{-1}$ coincide only when the condition \eqref{eq:s2match} holds.

Equations~\eqref{eq:Gmomentum}, \eqref{eq:Gentropy}, and \eqref{eq:firstlaw} define the tachydynamic variables of Galilean extremal surfaces. The quantity conjugate to the Bekenstein--Hawking entropy is the Galilei velocity $\vmakro$, while the corresponding conserved charge is the spatial momentum $P$.

In analogy with heat capacity, we define the \textit{tachydynamic capacity} by
\begin{align}
\mathcal C\equiv\frac{\dd P}{\dd\vmakro}
=\vmakro\frac{\dd S}{\dd\vmakro}\,.
\end{align}
One obtains
\begin{align}
\mathcal C=k\,\frac{|w'(X_0)|}{w''(X_0)}
\end{align}
which may have either sign.

\section{Galilean cosmologies}\label{sec:4}

The generic linear-dilaton solutions admit a natural homogeneous cosmological interpretation. The closed clock $\tau=\dd t$ provides a Galilean absolute time, while the Galilean Birkhoff Killing vector generates the one-dimensional homogeneous spatial leaves. The geometry can therefore be characterized by a spatial scale. We first derive its evolution equations and then determine the cosmological histories from the allowed dilaton region $M-w(X)>0$ and the behavior of the fields at its boundaries.

\subsection{Homogeneous 2d Galilean cosmologies}

A homogeneous one-dimensional spatial metric naturally gives a scale factor. From the solution \eqref{tau e solution}, we define
\begin{align}
a\equiv\left|e_\mu\xi^\mu\right|=e^{Q(X)}|\XP|\qquad
\qquad
h_{\mu\nu}\xi^\mu\xi^\nu=a^2\,.
\label{eq:Gscale}
\end{align}
Here, $a$ is invariant under Galilean boosts. Its normalization depends on the choice of Killing vector. 

The Casimir relation determines the spatial scale in terms of the dilaton,
\begin{align}
a^2=2e^{Q(X)}\left[M-w(X)\right]\,.
\label{eq:GscaleCasimir}
\end{align}
Reality requires $M-w(X)\geq0$. At a zero $\XP(t_i)=0$, provided $e^{Q(X_i)}$ is finite and non-vanishing,
\begin{align}
a(t_i)=0\qquad
\qquad
w(X_i)=M\,.
\end{align}
These zeros are degeneration loci of the homogeneous spatial scale $a(t)$ and coincide with the Galilean extremal locus discussed in Section~\ref{sec:3}.

On a regular branch with $\XP\neq0$, we define the Galilean Hubble rate by
\begin{align}
H_{\rm G}\equiv\frac{\dot a}{a}=K\,.
\end{align}
Here $K$ is the extrinsic curvature defined in Eq.~\eqref{eq:K}. 

The minimal torsion-free connection $\hat{\omega}$ is defined by $\dd e+\hat{\omega}\wedge\tau=0$ and satisfies $\hat{\omega}_\mu=K e_\mu$ in Eq.~\eqref{omegahat}. Using $K=H_{\rm G}$, $v=-\p_t$, and the curvature identity in Eq.~\eqref{electric limit}, the acceleration equation takes the form
\begin{align}
\frac{\ddot a}{a}&=\dot H_{\rm G}+H_{\rm G}^2=-v^\mu v^\nu\hat R_{\mu\nu}\,.
\end{align}
It has the one-dimensional Raychaudhuri form associated with the torsion-free connection $\hat{\omega}$. 
For $U=0$, the spatial torsion vanishes and the curvature is simply $-v^\mu v^\nu\hat R_{\mu\nu}=R/2$. The homogeneous equations then become
\begin{align}
H_{\rm G}=\frac{V}{\XP}\qquad\qquad \frac{\ddot a}{a}=\frac{R}{2}\,.
\end{align}

The spatial scale is stationary when $H_{\rm G}=0$ that requires
\begin{align}
V(X)=\frac12 U(X)\XP^2\,.
\end{align}
Unlike $\XP=0$, this condition does not have to involve a degeneration of the spatial metric.

\subsection{Cosmological branches and endpoints}
\label{sec:Gcosmo-global}

A cosmological branch covers an interval of $X$ where $M-w(X)>0$. Since $\dot X=-\XP\neq0$, the dilaton $X$ changes monotonically with time. The Galilean time interval is
\begin{align}
|\Delta t|=\left|\int_{X_1}^{X_2}\frac{e^{Q(X)/2}}{\sqrt{2[M-w(X)]}}\,\dd X\right|\,.
\label{eq:Gtimeextent}
\end{align}
Its convergence distinguishes finite-time endpoints from asymptotic boundaries.

For a simple root $X_i$, with $w(X_i)=M$, $w'(X_i)\neq0$, and finite non-vanishing $e^{Q(X_i)}$, the integrand in Eq.~\eqref{eq:Gtimeextent} behaves as $|X-X_i|^{-1/2}$, so one can reach it at a finite Galilean time $t_i$. Expanding Eqs.~\eqref{eq:Gtimeextent} and \eqref{eq:GscaleCasimir} near the root gives
\begin{align}
a(t)=2\pi\vmakro_i\,|t-t_i|+\mathcal O\bigl(|t-t_i|^3\bigr)\qquad\qquad |\dot a|\to2\pi\vmakro_i\,.
\label{eq:Gsimpleendpoint}
\end{align}
Thus $|H_{\rm G}|\sim|t-t_i|^{-1}$ near the endpoint, while the Galilean scalar curvature may remain finite here.

For a multiple root of finite order, assume that $e^{Q(X_*)}$ is finite and non-vanishing and, on an allowed side,
\begin{align}
M-w(X)\sim C|X-X_*|^n\qquad\qquad C>0\qquad\qquad n\geq2\,.
\end{align}
The time interval defined in~\eqref{eq:Gtimeextent} behaves as $\int\dd X\,|X-X_*|^{-n/2}$ and diverges. Hence $X_*$ is reached only at infinite Galilean time and defines an asymptotic boundary of the corresponding cosmological branch.

Curvature regularity must be analyzed separately. The location and existence of curvature singularities are model dependent and are not solely determined by $\XP=0$. The Galilei--Schwarzschild--Tangherlini and Galilei--Witten models shown in Sections~\ref{sec:GSTC} and \ref{sec:GWC} provide two explicit cases.

A regular branch bounded by two such simple roots $X_-<X_+$ has finite duration. Since $a>0$ in the interior and vanishes at both endpoints, the spatial scale has at least one interior maximum, where
\begin{align}
H_{\rm G}=0\,.
\end{align}
No uniqueness of this maximum follows for generic $U$ and $V$.

For an $S^2$ case, the matching condition \eqref{eq:s2match} implies
\begin{align}
\vmakro_-=\vmakro_+=\beta^{-1}
\end{align}
so Eq.~\eqref{eq:Gsimpleendpoint} gives equal endpoint slopes in magnitude. For a continuous $S^2$ family, it further has an $M$-independent $X_-+X_+=2X_*$ and reflection symmetry of $w$ about $X_*$. But this does not necessarily imply time-reflection symmetry of the cosmology.

A generic two simple roots branch does not have to satisfy the condition $\beta_{+}=\beta_{-}$ defined in \eqref{eq:s2match}; it then remains a finite-duration two-endpoint cosmology but lies outside the matched $S^2$ case considered above.

With a single simple endpoint, the opposite end of a branch is model dependent. It may lie at infinite Galilean time, as in the Galilei--CGHS model, or terminate at a curvature singularity, as in the Galilei--Schwarzschild and Galilei--Witten models.

\begin{figure}[t]
\centering
\includegraphics[width=\textwidth]{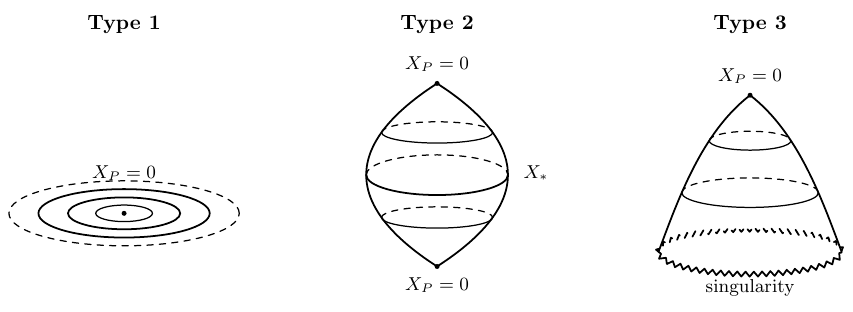}
\caption{Three representative Galilean cosmological histories: Type~1 (Galilei--CGHS) has a simple-root endpoint and an asymptotic end at infinite Galilean time; Type~2 (Galilei--JT) has two simple-root endpoints and finite duration; Type~3 (Galilei--Schwarzschild and Galilei--Witten) extends from a simple-root endpoint to a curvature singularity.}
\label{fig:galileicosmologytypes}
\end{figure}

The examples below realize three representative histories, shown schematically in Fig.~\ref{fig:galileicosmologytypes}.

\section{Examples}\label{sec:5}

In this Section, we study four Galilean dilaton models: Galilei--Schwarzschild--Tangherlini (GST), Galilei--Witten (GW), Galilei--Jackiw--Teitelboim (GJT), and Galilei--Callan--Gid\-dings--Harvey--Strominger (GCGHS). For each model, we give the vacuum solution, phase-space trajectories, cosmological evolution, and tachydynamic quantities. These models illustrate the three cosmological histories discussed in Section~\ref{sec:Gcosmo-global}: GCGHS (Type~1), GJT (Type~2), and GST and GW (Type~3).

The GST and GW cosmologies are related to the interiors of their relativistic counterparts, where the Killing vector becomes spacelike and bifurcation corresponds to a Galilean extremal surface. The GJT cosmology is related to an AdS$_2$ cosmological diamond, while GCGHS has vanishing curvature and denotes the flat-Galilean cosmology.

\subsection{Galilei--Schwarzschild--Tangherlini cosmologies}
\label{sec:GSTC}
\begin{figure}[htbp]
\centering
\makebox[\textwidth][c]{%
\begin{minipage}[t]{0.31\textwidth}
\centering
\includegraphics[width=\linewidth]{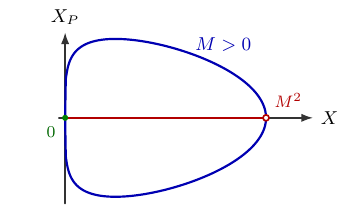}
\vspace{-0.6em}

{\small (a)}
\end{minipage}
\hspace{0.055\textwidth}
\begin{minipage}[t]{0.31\textwidth}
\centering
\includegraphics[width=\linewidth]{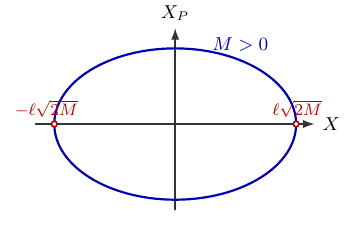}
\vspace{-0.6em}

{\small (b)}
\end{minipage}
}

\vspace{0.25em}

\makebox[\textwidth][c]{%
\begin{minipage}[t]{0.31\textwidth}
\centering
\includegraphics[width=\linewidth]{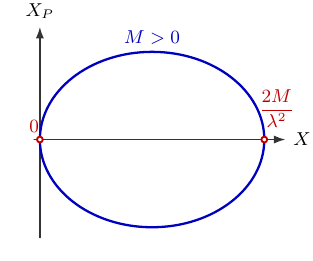}
\vspace{-0.6em}

{\small (c)}
\end{minipage}
\hspace{0.055\textwidth}
\begin{minipage}[t]{0.31\textwidth}
\centering
\includegraphics[width=\linewidth]{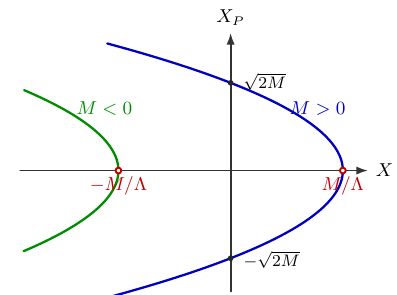}
\vspace{-0.6em}

{\small (d)}
\end{minipage}
}

\caption{Phase-space trajectories in the $(X,\XP)$ plane for (a) GST, (b) GJT, (c) GW, and (d) GCGHS cosmologies.}
\label{fig:phaseXXP}
\end{figure}

The GST model is the $D=4$ spherical-reduction theory discussed in Section~\ref{sec:2.3} \cite{Thomi:1984na,Benguria:1976in,Unruh:1976db,Berger:1972pg}. Setting $\lambda=2$, its potentials and associated functions are
\begin{align}
U_{\mathrm{GS}}(X)&=-\frac{1}{2X}&V_{\mathrm{GS}}(X)&=1
&e^{Q(X)}&=\frac{1}{2\sqrt{X}}&w(X)&=\sqrt{X}\,.
\end{align}
The Casimir conservation then gives
\begin{align}
\XP=\pm\sqrt{4\sqrt{X}\left(M-\sqrt{X}\right)}\,.
\end{align}
It requires $M>0$ and have $0<X<M^2$. The upper endpoint $X=M^2$ is a simple extremal surface, while $X=0$ is a curvature singularity. Each sign of $\XP$ therefore defines a Type~3  cosmological branch in the Fig.~\ref{fig:galileicosmologytypes}.
The corresponding geometric fields are
\begin{align}
e=\frac{\XP}{2\sqrt{X}}\,\dd r\qquad\qquad \omega=\frac{1}{\sqrt{X}}\left(1-\frac{M}{2\sqrt{X}}\right)\dd r\qquad\qquad R=\frac{2}{X}-\frac{2M}{X^{3/2}}\,.
\end{align}
The Galilean spatial metric is
\begin{align}
h_{\mu\nu}\dd x^\mu\dd x^\nu=\left(\frac{M}{\sqrt{X}}-1\right)\dd r^2\,.
\label{GS metric}
\end{align}
Thus the spatial scale vanishes at $X=M^2$ and diverges as $X\to0$. The curvature remains finite at the extremal endpoint, where $R\to0$, but diverges to $-\infty$ at the singular boundary. The two signs of $\XP$ are related by time reversal and describe distinct cosmological branches with the same extremal data at $X=M^2$. They form the two half-branches of the phase-space trajectory shown in Fig.~\ref{fig:phaseXXP}(a).

At the extremal endpoint $X_0=M^2$, the tachydynamic quantities of Section~\ref{sec:tachydynamics} are
\begin{align}
P&=\frac{k}{2\pi}M & \vmakro&=\frac{1}{4\pi M} & S&=kM^2 & \mathcal C&=-2kM^2\,.
\end{align}

\begin{figure}[htbp]
\centering
\includegraphics[width=0.80\textwidth]{"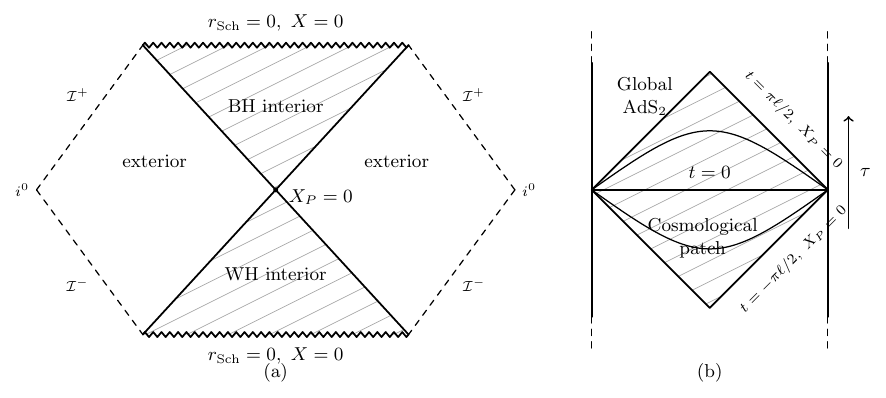"}
\caption{Relativistic geometries corresponding to the Galilean cosmologies. (a) Schwarzschild-type diagram for the GST and GW models. The black-hole and white-hole interiors correspond to two time-reversed Galilean branches. The Galilean extremal endpoint is associated with the horizon of the relativistic black hole. (b) AdS$_2$ cosmological patch associated with the GJT model, with extremal endpoints at $t=\pm\pi\ell/2$. Both panels depict Lorentzian geometries rather than Galilean spacetimes.}
\label{fig:cosmologicalstructures}
\end{figure}

The higher-dimensional origin gives this cosmology a direct interpretation as the Schwarzschild interior. For the $D=4$, $\lambda=2$, the dilaton $X$ and mass $M$ are related to the Schwarzschild radial and mass as (see~Eq.\eqref{eq:rescaleXlambda}) 
\begin{align}
X=\left(\frac{r_{\rm Sch}}{c}\right)^2\qquad \qquad M=\frac{2m_{\rm Sch}}{c}
\label{eq:GSdictionary}
\end{align}
where $r_{\rm Sch}$ and $m_{\rm Sch}$ denote the Schwarzschild areal radius and the mass parameter, respectively. Thus in standard Schwarzschild coordinates, one writes
\begin{align}
\dd s_{\rm Sch}^2=-c^2\left(1-\frac{2m_{\rm Sch}}{r_{\rm Sch}}\right)\dd t_{\rm Sch}^2+\left(1-\frac{2m_{\rm Sch}}{r_{\rm Sch}}\right)^{-1}\dd r_{\rm Sch}^2+r_{\rm Sch}^2\,\dd\Omega_2^2\,.
\end{align}
Using Eq.~\eqref{eq:GSdictionary}, the first $\dd t_{\rm Sch}^2$ term reproduces the Galilean spatial metric \eqref{GS metric} upon identifying $r=c\,t_{\rm Sch}$. Moreover, $0<X<M^2$ maps to $0<r_{\rm Sch}<2m_{\rm Sch}$, so the Galilean cosmological branch is inherited from the Schwarzschild interior. In this region $\partial_{t_{\rm Sch}}$ is spacelike, whereas the areal-radius direction is timelike; correspondingly, the Schwarzschild Killing direction becomes the Galilean spatial direction, while the Galilean time direction $t$ corresponds to the Schwarzschild radial direction $r_{\rm Sch}$, with $r_{\rm Sch}=c\sqrt{X}$. The extremal surface $X=M^2$ lies at the Schwarzschild horizon radius $r_{\rm Sch}=2m_{\rm Sch}$ and is associated with the bifurcation sphere. The two signs of $\XP$ are inherited from the black-hole and white-hole interiors and give the two time-reversed Galilean cosmological branches shown in Fig.~\ref{fig:cosmologicalstructures}(a).

\subsection{Galilei--Witten cosmology}
\label{sec:GWC}
The Lorentzian Witten black hole arises from the $SL(2,\mathbb{R})/U(1)$ coset construction~\cite{Witten:1991yr,Mandal:1991tz,Elitzur:1991cb} and admits a 2d dilaton-gravity description~\cite{Grumiller:2000wt,Grumiller:2021cwg}. Its Galilean counterpart is defined by
\begin{align}
U_{\mathrm{GW}}(X)=-\frac{1}{X}\qquad\qquad V_{\mathrm{GW}}(X)=\frac{\lambda^2}{2}X\qquad\qquad e^{Q(X)}=\frac{1}{X}\qquad\qquad w(X)=\frac{\lambda^2}{2}X
\end{align}
where $\lambda$ sets the curvature scale and satisfies $\lambda^2\propto1/\alpha^\prime$ in the string construction~\cite{Grumiller:2007ju}.
The Casimir relation gives
\begin{align}
\XP=\pm\sqrt{2MX-\lambda^2X^2}\,.
\end{align}
For $M>0$, the regular region is $0<X<2M/\lambda^2$, with the two signs of $\XP$ as shown in Fig.~\ref{fig:phaseXXP}(c).

All fields are solved as 
\begin{align}
X(t)&=\frac{2M}{\lambda^2}\sin^2\left(\frac{\lambda t}{2}\right) & \XP(t)&=-\frac{M}{\lambda}\sin(\lambda t)\\
e&=-\lambda\cot\left(\frac{\lambda t}{2}\right)\dd r & \omega&=\left(\lambda^2-\frac{M}{X}\right)\dd r\,.
\end{align}
The spatial metric and Galilean scalar curvature are
\begin{align}
h_{\mu\nu}\dd x^\mu\dd x^\nu&=\lambda^2\cot^2\left(\frac{\lambda t}{2}\right)\dd r^2 & R&=-\frac{2M}{X}=-\frac{2\lambda^2}{1-\cos(\lambda t)}\,.
\end{align}
At $t=\pi/\lambda$, one has $X=2M/\lambda^2$, $\XP=0$, and vanishing spatial scale, while the curvature remains finite, $R=-\lambda^2$. By contrast, $X\to0$ as $t\to0$ or $t\to2\pi/\lambda$, where $R\to-\infty$. The intervals $0<t<\pi/\lambda$ and $\pi/\lambda<t<2\pi/\lambda$ therefore give the two time-reversed Type~3 maximal branches shown in Fig~\ref{fig:galileicosmologytypes}. As in the GST example, they are inherited from the black-hole and white-hole interior sectors of the Lorentzian counterpart, as shown in Fig.~\ref{fig:cosmologicalstructures}(a).

At the regular extremal endpoint $X_0=2M/\lambda^2$, the tachydynamic quantities are
\begin{align}
P&=\frac{k}{2\pi}M & \vmakro&=\frac{\lambda^2}{4\pi} & S&=\frac{2kM}{\lambda^2}\,.
\end{align}
Since $\vmakro$ is independent of $M$, the tachydynamic capacity diverges, $\mathcal C\to\infty$.

\subsection{Galilei--Jackiw--Teitelboim model}
\label{sec:GJTC}
Jackiw--Teitelboim gravity \cite{Teitelboim:1983ux,Jackiw:1984je} provides the simplest constant-curvature benchmark in 2d dilaton gravity, see e.g.~\cite{Grumiller:2020elf,Gomis:2020wxp,Aviles:2022xyx}. Its Galilean counterpart is defined by
\begin{align}
U_{\mathrm{GJT}}(X)=0\qquad\qquad V_{\mathrm{GJT}}(X)=\frac{X}{\ell^2}\qquad\qquad e^{Q(X)}=1\qquad\qquad w(X)=\frac{X^2}{2\ell^2}\,.
\label{GJT potential}
\end{align}
The Casimir relation gives
\begin{align}
\XP=\pm\sqrt{2M-\frac{X^2}{\ell^2}}\,.
\end{align}
For $M>0$, the regular branch lies between the two simple roots
\begin{align}
X_\pm=\pm\ell\sqrt{2M}\,.
\end{align}
The phase-space orbit is the closed ellipse shown in Fig.~\ref{fig:phaseXXP}(b). 

Since $w(X)$ is reflection symmetric about $X_*=0$ and
\begin{align}
|w'(X_+)|=|w'(X_-)|=\frac{\sqrt{2M}}{\ell}
\end{align}
which admits a global period
\begin{align}
\beta=\frac{2\pi\ell}{\sqrt{2M}}\,.
\end{align}
It gives the closed Type~2 cosmology of Fig.~\ref{fig:galileicosmologytypes}.

Choosing $t=0$ at the midpoint $X=0$, the solution is
\begin{align}
X(t)&=\sqrt{2M\ell^2}\sin\frac{t}{\ell} & \XP(t)&=-\sqrt{2M}\cos\frac{t}{\ell}\label{eq:XJT}\\
e&=\XP(t)\,\dd r & \omega&=\frac{X(t)}{\ell^2}\,\dd r\,.
\end{align}
The spatial metric and Galilean scalar curvature are
\begin{align}
h_{\mu\nu}\dd x^\mu\dd x^\nu&=2M\cos^2\frac{t}{\ell}\,\dd r^2 & R&=-\frac{2}{\ell^2}\,.
\end{align}
The maximal cosmological history has finite duration,
\begin{align}
-\frac{\pi\ell}{2}<t<\frac{\pi\ell}{2}
\end{align}
with vanishing spatial scale $a(t)$ at both two extremal endpoints and a unique maximum at $t=0$. The curvatures at both two endpoints are regular. The associated Lorentzian JT geometry is the cosmological patch of global AdS$_2$ shown in Fig.~\ref{fig:cosmologicalstructures}(b).

Hence, the allowed interval of $X$ in this GJT model includes negative dilaton values. From the point of view of interpreting dilaton $X$ as transverse area or an inverse gravitational coupling~\cite{Grumiller:2002nm}, the positivity of $X$ is natural. But it is not required by the classical JT equations of motion. Such sign-changing profiles already occur in the general Lorentzian JT solutions (see Section 2.1 of Ref.~\cite{Harlow:2018tqv}), also a closed-universe solution with the same sinusoidal $X(t)$ in Eq.\eqref{eq:XJT} in the Ref.~\cite{Usatyuk:2024mzs}. We therefore retain the full classical solution, including $X<0$,  which leads to the Type~2 closed universe interpretation in Fig~\ref{fig:galileicosmologytypes}. But imposing $X>0$ would truncate the cosmological history at the regular slice $t=0$, see Fig~\ref{fig:cosmologicalstructures}.

Using the $X_*=0$ as the entropy reference defined in Eq.~\eqref{eq:Gentropy}, the two extremal surfaces carry same G-tachydynamic data
\begin{align}
P&=\frac{k}{2\pi}M &
\vmakro_+=\vmakro_-&=\frac{\sqrt{2M}}{2\pi\ell} &
S_+=S_-&=k\ell\sqrt{2M} &
\mathcal C_+=\mathcal C_-&=k\ell\sqrt{2M}\,.
\end{align}

\subsection{Galilei--CGHS model}

The Callan--Giddings--Harvey--Strominger model was introduced as an exactly solvable setting for 2d black-hole evaporation coupled to conformal matter~\cite{Callan:1992rs}. In the Galilean theory it is specified by
\begin{align}
U_{\mathrm{GCGHS}}(X)=0\qquad\qquad V_{\mathrm{GCGHS}}(X)=\Lambda\qquad\qquad e^{Q(X)}=1\qquad\qquad w(X)=\Lambda X
\end{align}
with $\Lambda\neq0$.
The Casimir relation becomes
\begin{align}
\XP=\pm\sqrt{2\left(M-\Lambda X\right)}\,.
\end{align}
There is a single simple extremal root,
\begin{align}
X_0=\frac{M}{\Lambda}
\end{align}
for either sign of $M$. The two signs of $\XP$ give the time-reversed branches of the parabolic phase-space curve shown in Fig.~\ref{fig:phaseXXP}(d).

Placing the extremal endpoint at $t=0$, the solution reads
\begin{align}
X(t)=\frac{M}{\Lambda}-\frac{\Lambda}{2}t^2\qquad\qquad \XP(t)=\Lambda t\,,\qquad e=\Lambda t\,\dd r\qquad\qquad \omega=\Lambda\,\dd r\,. \label{eq:GCGHS X}
\end{align}
The geometry is flat,
\begin{align}
h_{\mu\nu}\dd x^\mu\dd x^\nu=\Lambda^2t^2\,\dd r^2\qquad\qquad R=0\,.
\end{align}
The spatial scale vanishes at $t=0$ and grows linearly as $|t|\to\infty$. Each sign branch therefore extends from the extremal starting-points to an asymptotic end reached only at infinite Galilean time, as the flat/open Type~1 cosmology in Fig.~\ref{fig:galileicosmologytypes}.

Here, we choose $\Lambda>0$ as in the standard CGHS model~\cite{Callan:1992rs}. But the solution~\eqref{eq:GCGHS X} shows that $X$ is unbounded below for either sign of the mass $M$, and crosses $X=0$ in the $M>0$ case; see the phase diagram in Fig.~\ref{fig:phaseXXP}(d). However, in the standard CGHS theory~\cite{Callan:1992rs}, $X=e^{-2\phi}>0$ for real $\phi$, and hence $X\rightarrow 0^{+}$ corresponds to the strong-coupling limit.
Additionally, the Weyl transformation relating the original CGHS model to its equivalent constant-potential formulation is singular at $X=0$~\cite{Grumiller:2002nm}.
Imposing $X>0$ would end the phase orbit at a finite Galilean time $t$. The continuation to $X<0$ cannot retain the original CGHS interpretation. Nevertheless, $X>0$ need not be imposed once we treat this linear-constant model as an intrinsic Galilean theory. We therefore retain the continuation to the negative-dilaton regime, which leads to the Type~1 cosmology picture shown in Fig.~\ref{fig:galileicosmologytypes}. This extended solution, hence, no longer admits either the $X\sim e^{-2\phi}$ or the interpretation as an effective inverse Newton constant. As discussed, a  similar situation also occurs in the GJT case shown in the previous Subsection.

At the extremal surface,
\begin{align}
P&=\frac{k}{2\pi}M & \vmakro&=\frac{|\Lambda|}{2\pi}\,.
\end{align}
For a fixed entropy reference $X_{\rm ref}$ on the chosen branch,
\begin{align}
S=k\,\operatorname{sgn}(\Lambda)\left(\frac{M}{\Lambda}-X_{\rm ref}\right)\,.
\end{align}
In particular, for $\Lambda>0, M>0$, we set $X_{\rm ref}=0$
\begin{align}
P=\frac{k}{2\pi}M\qquad\qquad \vmakro=\frac{\Lambda}{2\pi}\qquad\qquad S=\frac{k}{\Lambda}M\,.
\end{align}
Since $\vmakro$ is independent of $M$, the tachydynamic capacity diverges,
\begin{align}
\mathcal C\to\infty\,.
\end{align}

\section{Conclusions and outlook}

In this work, we have constructed and investigated the most general 2d Galilean dilaton gravity theory with a potential quadratic in the auxiliary field. Starting from its Poisson sigma model formulation, we derived the corresponding second-order action and established its relation to the large-$c$ limit of relativistic dilaton gravity. We further demonstrated that, under an appropriate scaling of the higher-dimensional gravitational coupling and radial coordinate, the Galilean limit of spherically symmetric Einstein gravity reproduces the same 2d theory. In this specific setting, spherical reduction and the Galilean limit therefore commute, up to boundary terms.

We constructed the general classical solutions and established a Galilean version of the Birkhoff theorem. In contrast to conventional Lorentzian black-hole geometries, the generic solutions admit a spatial Killing vector and naturally describe homogeneous Galilean cosmologies. Their evolution is governed by the conserved Casimir and the dilaton potential, allowing for a classification of cosmological branches according to their endpoint structure.

A central feature of these solutions is the locus $X_{\textrm{\tiny P}}=0$, which we identified as a Galilean extremal surface. It is simultaneously a fixed point of Galilean boosts in the Poisson sigma model target space, a stationary point of the dilaton along the absolute-time direction, and, under suitable regularity conditions, an endpoint of the homogeneous spatial geometry. Although the Galilean spatial structure degenerates there, the underlying manifold can remain smooth.

We also developed the associated Galilean tachydynamics. Using the covariant phase-space formalism, we identified the conserved spatial momentum and derived a first law relating its variation to the entropy variation through the Galilei velocity. This provides a Galilean counterpart of black-hole thermodynamics in which the role of temperature is played by a characteristic velocity.

The explicit models considered illustrate different cosmological histories. In particular, the Galilei--Schwarzschild--Tangherlini and Galilei--Witten models connect the Galilean cosmological branches to the interiors of their relativistic black-hole counterparts. Their Galilean extremal surfaces are associated with the relativistic bifurcation surfaces. The Galilei--Jackiw--Teitelboim model provides a regular closed cosmology, while the Galilei--CGHS model illustrates an open cosmological branch with an asymptotic endpoint.

Several questions remain open. It would be interesting to investigate the quantum theory of Galilean dilaton gravity, including its partition functions and the quantum interpretation of Galilean extremal surfaces. Another natural direction is to identify the boundary degrees of freedom associated with these surfaces and establish their relation to Carrollian holography and asymptotically flat holography. Finally, extending the present construction beyond two dimensions may provide further insight into the relation between Galilean cosmologies, black-hole interiors and non-Lorentzian gravitational dynamics.


\acknowledgments

We thank Florian Ecker and Jelle Hartong for discussions.

This work was supported by the Austrian Science Fund (FWF) [Grants DOI: \href{https://www.fwf.ac.at/en/research-radar/10.55776/P36619}{10.55776/P36619}, \href{https://www.fwf.ac.at/en/research-radar/10.55776/PAT1871425}{10.55776/PAT1871425}]. K.-Y. Zhang acknowledges financial support from the China Scholarship Council (CSC).


\appendix
\section{Galilean symmetry}\label{app:A}
\label{Galilean algebra}

Galilean symmetry emerges as a non-relativistic limit of Poincar\'e symmetry. In this Appendix, we review the Galilei algebra in arbitrary dimensions from the perspective of \.In\"on\"u--Wigner contractions.

We begin with the $(d+1)$-dimensional Poincar\'e algebra $\mathfrak{iso}(1,d)$. Its generators consist of the time translation $H$, spatial translations $P_a$, rotations $J_{ab}$, and boosts $B_a$, where the spatial indices run as $a, b = 1, \dots, d$. The rotation generators $J_{ab}$ form an antisymmetric tensor, $J_{ab} = -J_{ba}$. The non-vanishing commutation relations read

\begin{align}
    [J_{ab},J_{cd}] &= 4\delta_{[c[a}J_{b]d]}
    & [J_{ab},P_c]&=2\delta_{c[a}P_{b]}
    & [J_{ab},B_c]&=2\delta_{c[a}B_{b]} \\
    [B_a,B_b] &= -\frac{1}{c^2}J_{ab}
    & [B_a,P_b]&=\frac{1}{c^2}\delta_{ab}H
    & [B_a,H]&=P_a\,.
\end{align}
A realization of these generators in terms of differential operators is given by
\begin{equation}
H =\partial_t \qquad\qquad P_a =\partial_a \qquad\qquad
J_{ab} =x_a\,\partial_b - x_b\,\partial_a \qquad\qquad B_a =-t\,\partial_a + \frac{1}{c^2}\,x_a\,\partial_t\,.
\end{equation}
Note that we have rescaled the Lorentzian boost generator to give it units of inverse velocity. This is useful for the Galilean limit obtained via an \.In\"on\"u--Wigner contraction \cite{Inonu:1953sp}, i.e., the singular limit $c \to \infty$.

In the $c\to\infty$ limit, the non-contracted subalgebra is $\mathfrak{iso}(d)\oplus\mathbb{R}_H$, generated by rotations and translations, $\{J_{ab}, P_a, H\}$, while the contracted generators correspond to the boosts $B_a$.

We define the Galilean boost generators $G_a$ as a limit of the Lorentzian ones,
\begin{equation}
G_a \coloneqq \lim_{c\rightarrow \infty} B_a =-t\,\partial_a\,.
\end{equation}
The corresponding commutators are obtained by taking the limit of the Poincar\'e commutation relations. The resulting Galilei algebra $\mathfrak{gal}(d+1)$ is generated by $\{J_{ab}, P_a, G_a, H\}$ with the non-vanishing commutators
\eq{
[J_{ab},\,J_{cd}] =4\delta_{[c[a} J_{b]d]} 
\qquad [J_{ab},\,P_c] =2\delta_{c[a} P_{b]} 
\qquad [J_{ab},\,G_c] =2\delta_{c[a} G_{b]} 
\qquad [G_a,\,H] = P_a\,. 
}{gal(1,d)}

Exponentiating the boost generators $G_a$ with velocity parameters $v^a$ leads to the finite Galilean boost transformations. In contrast to the Lorentzian case, where boosts mix temporal and spatial components, the action of the group element $g(\vec{v}) = \exp(v^a G_a)$ on the spacetime coordinates $(t, x^a)$ is given by
\begin{equation}
t^\prime = t \qquad\qquad x^{\prime\,a} = x^a - v^a\, t\,.
\end{equation}
This transformation explicitly reflects the absolute nature of time in the non-relativistic regime. Furthermore, since the boost generators form an abelian subalgebra, $[G_a, G_b] = 0$, successive boosts commute. As a result, the composition of two boosts leads to a simple linear addition of velocities, $\vec{v}_{12} = \vec{v}_1 + \vec{v}_2$, and does not give rise to a Wigner rotation or Thomas precession.

In a 2d context, spatial translations $P$ commute with all generators and thus constitute a central element. It is worth noting the isomorphism $\mathfrak{gal}(2) \simeq \mathfrak{carr}(2)$, which is realized via the map $H_{\text{gal}} \leftrightarrow P_{\text{carr}}$, $P_{\text{gal}} \leftrightarrow H_{\text{carr}}$, and $G_{\text{gal}} \leftrightarrow C_{\text{carr}}$.

In contrast to the Carroll algebra $\mathfrak{carr}(1+d)$, for $d > 1$, the Galilei algebra $\mathfrak{gal}(d+1)$ admits a non-trivial one-dimensional central extension. This extension is characterized by a 2-cocycle which modifies the commutator between boosts and spatial translations. The resulting Bargmann algebra $\mathfrak{barg}(d+1)$ is generated by $\{J_{ab}, P_a, G_a, H, M\}$, where $M$ is the central element. The non-vanishing commutators are given by \eqref{gal(1,d)} and additionally
\begin{equation}
\textrm{Bargmann\;algebra:\;Eqs.~\eqref{gal(1,d)}\;and} \qquad [G_a, P_b] =\delta_{ab}\, M
\end{equation}
The consistency of the Bargmann algebra is guaranteed by the Jacobi identities. The generator $M$ is interpreted as mass and the fact that it is central means that mass is conserved and we have a superselection rule --- a Bargmann universe cannot be in a superposition of different mass eigenstates. The Bargmann algebra has found several nonrelativistic applications, see e.g.~\cite{Duval:1984cj,Andringa:2010it,Bergshoeff:2016lwr}. In the present work, we exclusively employ the (non-extended) Galilei algebra.

\section{Galilean gravity}\label{app:B}

In this Appendix, we outline the fundamental features of Galilean gravity, with a particular focus on the 2d case. For a comprehensive review, see for instance \cite{Hartong:2022lsy,Duval:2014uoa}.

\subsection{Flat Galilean manifold}
The intrinsic geometric description of non-relativistic spacetimes traces back to Cartan \cite{Cartan:1923zea,Cartan:1924yea}. This framework was further developed by Trautman and K\"unzle to characterize Galilean geometry through independent degenerate metric structures rather than as a limit of GR \cite{Kuenzle:1972zw,Trautman1963}.

An intrinsic definition of a Galilean spacetime is provided by the quadruple $(G, \ggtna{},\tau, \Pi)$, as adopted in \cite{Duval:2014uoa}. Here, $G$ denotes a smooth $(d+1)$-dimensional manifold, $\Pi^{\m\n}$ is a symmetric, contravariant, positive-semidefinite tensor field, and $\tau_\mu$ is the clock 1-form. The 1-form $\tau$ is nowhere vanishing and spans the kernel of the cometric $\Pi$, satisfying $\Pi^{\mu\nu}\tau_\nu = 0$. Additionally, the geometry is equipped with a Galilean connection $\ggtna{_\m}$ that satisfies the two compatibility conditions
\begin{equation}
    \ggtna{_\m}\Pi^{\r\s}=0\qquad\qquad \ggtna{_\m}\tau_{\n}=0\,.
\end{equation}

The flat Galilean manifold is defined by choosing the standard flat
compatible connection, with the remaining connection ambiguity set to
zero. In Cartesian coordinates the resulting structure is
\begin{equation}
    G = \mathbb{R} \times \mathbb{R}^{d} \qquad\qquad
    \Pi = \delta^{ij}\p_i \p_j \qquad\qquad
    \tau = \dd t \qquad\qquad
    \ggtGa{^i_{jk}} = 0 
\end{equation}
where $t$ denotes the time coordinate and indices $i,j$ run over spatial dimensions. The automorphisms of such a manifold are generated by vector fields $\xi^\m$ that preserve the entire geometric structure. These conditions are expressed as
\begin{subeqs}
         \mathcal{L}_{\xi} \Pi^{\m\n} &= 0 &
        \mathcal{L}_{\xi} \tau_\m &= 0 \label{killing b} \\
        \delta_{\xi}\ggtGa{^\r_{\m\n}} &= \p_\m\p_\n \xi^\r = 0 \,. &&\label{auto flat}
\end{subeqs}
   
 The constraint on the connection \eqref{auto flat} restricts $\xi^\mu$ to be an affine function of the coordinates, while the preservation of the cometric and time 1-form fixes the coefficients. The resulting general solution is given by
\begin{equation}
    \xi_{\mathrm{auto}} = \epsilon \p_t + \left( \omega^i{}_j x^j + v^i t + a^i \right) \p_i 
\end{equation}
where the constant parameters correspond to the generators of the $(d+1)$-dimensional Galilean algebra $\mathfrak{gal}(1,d)$: $\epsilon$ generates time translations, $a^i$ spatial translations, $v^i$ Galilean boosts, and the antisymmetric matrix $\omega_{ij}$ generates spatial rotations. Thus, the group of affine automorphisms of the flat Galilean manifold is identified as the Galilean group. 

In the absence of a non-degenerate metric tensor, one may consider a broader class of symmetries that preserve only the metric structure $(\tau, \Pi)$ but not necessarily the connection. These are the Galilean Killing vectors, defined solely by conditions \eqref{killing b}. The most general solution to this relaxed set of equations is
\eq{
    \xi_{\mathrm{k}}=\epsilon \p_t + \left( \omega^i{}_j x^j + f^i(t) \right) \p_i 
}{flat killing}
where $f^i(t)$ are $d$ arbitrary functions of time. The presence of these infinitely many Killing vector fields, of the form $\xi_f=f^i(t)\p_i$, constitutes an analogue of the Carrollian supertranslations $\xi_f=f(x^i)\p_t$ which preserve the Carrollian structure \cite{Ecker:2023uwm,Henneaux:2021yzg}.

\subsection[Two dimensional Galilean gravity via gauging \texorpdfstring{$\mathfrak{gal}(2)$}{gal(2)}]{Two dimensional Galilean gravity via gauging \texorpdfstring{$\boldsymbol{\mathfrak{gal}(2)}$}{gal(2)}}
An approach to construct a gravity theory is the gauge procedure, see e.g.~\cite{Hartong:2016yrf,Andringa:2010it,Campoleoni:2022ebj,Bergshoeff:2017btm,Matulich:2019cdo}.
The construction of 2d Galilean gravity starts with the definition of the connection 1-form $\mathcal{A}_{\mu}$ taking values in the Galilean algebra $\mathfrak{gal}(2)$. Spanned by the generators of time translations ($H$), space translations ($P$), and Galilean boosts ($G$), the gauge connection is parameterized as
\begin{align}
    \mathcal{A}_{\mu} & =\tau_{\mu} H+e_{\mu} P+\omega_{\mu} G\,.
\end{align}
The algebraic structure is characterized by the single non-vanishing commutation relation
\begin{equation}
    [G, H] = P\,.
\end{equation}
Under a gauge transformation generated by the parameter $\lambda = \lambda_{H} H + \lambda_{P} P + \lambda_{G} G$, the connection transforms according to $\delta_{\lambda}\mathcal{A}_{\mu} = \mathcal{D}_{\mu} \lambda = \partial_{\mu} \lambda + [\mathcal{A}_{\mu}, \lambda]$.
Consequently, the component fields transform as
\begin{align}
    \delta_{\lambda} \tau_{\mu} & = \partial_{\mu} \lambda_{H}  &
    \delta_{\lambda} e_{\mu} & = \partial_{\mu} \lambda_{P} + \omega_{\mu}\lambda_{H} - \lambda_{G} \tau_{\mu}  &
    \delta_{\lambda} \omega_{\mu} & = \partial_{\mu} \lambda_{G}\,. 
\end{align}

To relate the gauge fields to the spacetime geometry, we introduce a background affine connection $\ggtna{_\mu}$ and impose the vielbein postulates
\begin{align}
    D_{\mu} \tau_{\nu} & \equiv \ggtna{_\mu} \tau_{\nu} = 0 \label{postulate tau} \\
    D_{\mu} e_{\nu} & \equiv \ggtna{_\mu} e_{\nu} + \omega_{\mu} \tau_{\nu} = 0\,. \label{postulate e}
\end{align}
The first postulate, Eq.~\eqref{postulate tau}, ensures the absolute nature of the time foliation, while the second, Eq.~\eqref{postulate e}, couples the spatial vielbein to the boost gauge field $\omega_\mu$.

We define the inverse basis vectors $(v^\mu, \theta^\mu)$ via the orthogonality and normalization relations
\begin{align}
    v^{\mu} \tau_{\mu} & = -1 & \theta^{\mu} e_{\mu} & = 1  &
    v^{\mu} e_{\mu} & = 0 & \tau_{\mu} \theta^{\mu} & = 0\,. \label{eq:def theta}
\end{align}
The completeness of the basis $\{\tau_{\mu}, e_{\nu}\}$ implies the resolution of the identity
\begin{align}
    \delta_{\nu}^{\mu} & = -v^{\mu} \tau_{\nu} + \theta^{\mu} e_{\nu}\,. \label{completeness relation}
\end{align}
This structure permits the definition of a degenerate spatial metric $h_{\mu\nu} = e_{\mu}e_{\nu}$ and its inverse cometric $h^{\mu\nu} = \theta^{\mu}\theta^{\nu}$.

Demanding consistency between the gauge transformations and the vielbein postulates fixes the transformation rules and covariant derivatives of the inverse basis. Specifically, under a Galilean boost parameterized by $\lambda_G$, the fields transform according to
\begin{align}
    \delta_{\lambda_{G}} \tau_{\mu} & = 0 & \delta_{\lambda_{G}} v^{\mu} & = -\lambda_{G} \theta^{\mu} \label{eq:delta v} \\
    \delta_{\lambda_{G}} e_{\mu} & = -\lambda_{G} \tau_{\mu} & \delta_{\lambda_{G}} \theta^{\mu} & = 0  \\
    \delta_{\lambda_{G}} \omega_{\mu} & = \partial_{\mu} \lambda_{G}\,. & & 
\end{align}
Similarly, the covariant derivatives of the inverse basis vectors are given by
\begin{align}
    \ggtna{_\mu} \tau_{\nu} & = 0 & \ggtna{_\nu} v^{\mu} & = -\omega_{\nu} \theta^{\mu} \label{postulate tau_inv} \\
    \ggtna{_\mu} e_{\nu} & = -\omega_{\mu} \tau_{\nu} & \ggtna{_\nu} \theta^{\mu} & = 0\,. \label{postulate e_inv}
\end{align}

The affine connection $\Gamma\indices{^\rho_{\mu\nu}}$ can be derived in terms of the gauge fields by expanding the vielbein postulates \eqref{postulate tau} and \eqref{postulate e} as
\begin{align}
    \partial_{\mu} \tau_{\nu} &= \Gamma\indices{^\s_{\mu\nu}} \tau_{\s} \label{Gamma 1} \\
    \partial_{\mu} e_{\nu} &= \Gamma\indices{^\s_{\mu\nu}} e_{\s} - \omega_{\mu} \tau_{\nu}\,. \label{Gamma 2}
\end{align}
Contracting Eq.~\eqref{Gamma 1} with $-v^\rho$ and Eq.~\eqref{Gamma 2} with $\theta^\rho$, and summing the results using the completeness relation \eqref{completeness relation}, we obtain
\begin{align}
    \Gamma\indices{^\rho_{\mu\nu}} & = -v^{\rho}(\partial_\mu \tau_\nu) + \theta^\rho (\partial_\mu e_\nu + \omega_\mu \tau_\nu)\,.
    \label{Gamma explicit}
\end{align}

The gauge curvature 2-form, $R \equiv \dd\mathcal{A} + [\mathcal{A} \wedge \mathcal{A}]$, decomposes into components associated with each generator
\begin{align}
    R[H] & = \dd\tau  &
    R[P] & = \dd e + \omega \wedge \tau  &
    R[G] & = \dd\omega\,.
\end{align}
The affine torsion tensor, $T\indices{^\rho_{\mu\nu}} \equiv \Gamma\indices{^\rho_{\mu\nu}} - \Gamma\indices{^\rho_{\nu\mu}}$, is directly related to these gauge curvatures. Substituting the explicit form of the connection \eqref{Gamma explicit}, we find
\begin{align}
    T\indices{^\rho_{\mu\nu}} & = -v^{\rho}R[H]_{\mu\nu} + \theta^{\rho}R[P]_{\mu\nu}\,.
\end{align}
Finally, using the inverse-vielbein postulates \eqref{postulate tau_inv}--\eqref{postulate e_inv} and the Riemann-tensor convention \eqref{Riemann D}, we obtain
\begin{align}
    R_{\mu\nu\sigma}{}^\rho \theta^\sigma &= 0
    &
    R_{\mu\nu\sigma}{}^\rho v^\sigma &= \theta^\rho R[G]_{\mu\nu}\,.
\end{align}
Using the completeness relation \eqref{completeness relation}, the Ricci tensor is therefore
\begin{align}
    R_{\mu\nu}
    :=R_{\mu\rho\nu}{}^\rho
    =-\tau_\nu\theta^\rho R[G]_{\mu\rho}\,. \label{2Ricci}
\end{align}

\section{Large-\texorpdfstring{$\boldsymbol{c}$}{c} expansion of Einstein gravity}\label{app:C}
\subsection[A \texorpdfstring{$1+d$}{1+d} decomposition of the Lorentzian metric]{A \texorpdfstring{$\boldsymbol{1+d}$}{1+d} decomposition of the Lorentzian metric}
The pre-non-relativistic (PNR) and pre-ultra-local (PUL) parameterizations, as proposed in \cite{Hansen:2021fxi,Hansen:2020pqs}, provide powerful frameworks for expanding the Einstein-Hilbert action in the large-$c$ (Galilean) and small-$c$ (Carrollian) regimes, respectively. These two dual parameterization schemes rely on a common local decomposition of the metric for a $(d+1)$-dimensional Lorentzian manifold with signature $(-1,1\dots,1)$. Specifically, the metric is expressed in terms of local variables, distinguishing a temporal structure from the spatial metric
\begin{subeqs}
    g_{\m\n}& = -c^2 T_\m T_\n + \Pi_{\m\n}\label{1+d a}\\
    g^{\m\n}&=-\frac{1}{c^2} V^\m V^\n +\Pi^{\m\n}\,.\label{1+d b}
\end{subeqs}
For the large-$c$ expansion, we call $V^\m,T_\m,\Pi^{\m\n}, \Pi_{\m\n}$ the pre-nonrelativistic variables.
These variables satisfy the following orthogonality and completeness relations
\begin{subeqs}
T^\m V_\m&=-1 & V^\m \Pi_{\m\n}&=0=T_\m \Pi^{\m\n}\label{pre g a}\\
\ts{\delta}{^\m_\n}&=-V^\m T_\n+\Pi^{\m\r}\Pi_{\r\n}\,. && \label{pre g b}
\end{subeqs}
By introducing local spatial vielbeins defined by
\begin{eqs}
    \Pi_{\m\n}&=\delta_{ab}\tcE{^a_\m}\tcE{^b_\n}&\quad \Pi^{\m\n}&=\delta^{ab}\tbth{_a^\m}\tbth{_b^\n}\\
   \ts{\delta}{^a_b}&=\tcE{^a_\m}\tbth{_b^\m}&\quad \ts{\delta}{^\m_\n}+V^\m V_\n&=\tcE{^a_\m}\tbth{_a^\n}
\end{eqs}where lowercase Latin indices $a,b,c\dots$ denote spatial directions, we can organize these local variables into two generalized vielbein multiplets
\begin{align}
    \tE{^A_\m}=\big(cT_\m,\tcE{^a_\m}\big)\qquad
    \tbth{_A^\m}=\left(-\frac{1}{c}V^\m,\tbth{_a^\m}\right)\qquad
    \tE{^A_\m}\tbth{_A^\n}=\delta^\n_\m\qquad
    \tE{^A_\m}\tbth{_B^\m}=\delta^A_B\,.
    \label{c decom 2}
\end{align}
Here, the capital index $A=(0,a)$ encompasses both temporal and spatial components. Consequently, the metric decomposition in Eqs. \eqref{1+d a} and \eqref{1+d b} can be recast as
\begin{eqs}
    g_{\m\n}=\eta_{AB}\tE{^A_\m}\tE{^B_\n}\quad
    g^{\m\n}=\eta^{AB}\tbth{_A^\m}\tbth{_B^\n}\quad
    \eta_{AB}=g_{\m\n}\tbth{_A^\m}\tbth{_B^\n}\quad
    \eta^{AB}=g^{\m\n}\tE{^A_\m}\tE{^B_\n}\,.
\end{eqs}
It is worth noting that a standard Lorentzian boost does not preserve the parameterization defined in Eq. \eqref{c decom 2}, as such a transformation mixes the temporal components ($T_\m, V^\m$) with the spatial vielbeins ($\cE^a_\m, \Theta_a^\m$).

\subsection{Determination of Galilean compatible connection via first-order formalism}
In the Cartan formalism, the vielbein postulate establishes the relationship between the spin connection $\Omega_{AB}$ and the manifold connection
\begin{align}
    \nabla_\mu \tE{^A_\n} = -\tom{_\m^A_B} \tE{^B_\n}\,.\label{pos1}
\end{align}
Similarly, the covariant derivative of the inverse vielbein is given by
\begin{align}
    \nabla_\m\tbth{_A^\n}=+\tom{_\m^B_A}\tbth{_B^\n}\,.\label{pos2}
\end{align}
Using the completeness relation for $E_\m^A$, the Christoffel symbols associated with the connection $\nabla_\m$ can be expressed in terms of $E$ and $\Omega$ as
\begin{eqs}
    \tGa{^\r_{\m\n}} &= \tbth{_A^\r} \p_\m \tE{^A_\n} + \tbth{_A^\r} \tom{_\m^A_B} \tE{^B_\n} \\
    &=-\frac{1}{c}V^\r\sb{c\p_\m T_\n+\tom{_\m^0_a}\tcE{^a_\n}}+\tbth{_a^\r}\sb{\p_\m \tcE{^a_\n}+c\tom{_\m^a_0}T_\n+\tom{_\m^a_b}\tcE{^b_\n}}\label{christoffel}
\end{eqs}
In standard Riemannian geometry, the metric compatibility and torsion-free conditions provide a sufficient number of constraints to uniquely determine the spin connection $\tom{_\m^A_B}$ in terms of the vielbein forms $\tE{^A_\m}$. In that context, the Christoffel symbols \eqref{christoffel} are uniquely determined by $(T_\m,\tcE{^a_\m})$, or equivalently by the metric $g_{\m\n}=-c^2T_\m T_\n+\delta_{ab}\tcE{^a_\m}\tcE{^b_\n}$. This result is encapsulated in the Levi-Civita theorem, and the resulting connection satisfying $\nabla_\m g_{\r\s}=0$ is termed the Levi-Civita connection.

However, this uniqueness does not hold when the compatibility conditions on $\nabla_{\m}$ are modified, as the torsion-free conditions cannot be maximally imposed, as will be demonstrated in the following Section.

We now introduce the pre-Galilean connection, denoted by $\tilde{G}$, which satisfies the two compatibility conditions
\begin{subeqs}
    \gtna{_\m}T_{\n}=0 \qquad\qquad \gtna{_\r}\Pi^{\m\n}=0\,.
\end{subeqs}
It is important to note that the second condition is not equivalent to $\gtna{_\r} \tbth{_a^\m}=0$ for general dimensions $d\ge2$; however, they are equivalent in 2d spacetime. Applying the vielbein postulates \eqref{pos1} and \eqref{pos2} yields
\begin{subeqs}
    \gtom{^0_0}=0=\gtom{^0_a} \qquad\qquad \gtom{_a_b}=-\gtom{_b_a}\,.
\end{subeqs}
It is convenient to denote
\begin{eqs}
    \gtom{^a}:= \gtom{^a_0}
\end{eqs}

The torsion $2$-form in the Cartan formalism is defined by
\begin{align}
    T^A \coloneqq DE^A= \dd E^A + \omega^A{}_B \wedge E^B.
\end{align}
The individual components of the torsion $2$-form associated with the Galilean spin connection are listed as follows
\begin{subeqs}
    T^{0}&=c\dd T\\
    T_{a}(V, \Theta_b)&=\dd \mathcal{E}_a(V, \Theta_b)+\omega_{ab}(V)+c\omega_{a}(\Theta_b)\\
    T_{a}(\Theta_b, \Theta_c)&=\dd\mathcal{E}_a(\Theta_b, \Theta_c)-2 \omega_{a[b}(\Theta_{c]})\,.
\end{subeqs}
The first identity indicates that the pre-Galilean connection generally possesses a non-zero intrinsic torsion proportional to the temporal $2$-form $\dd T$. We can rearrange the second and third identities to express $\omega_{ab}$ in terms of $\omega_{a}, T^a$, and $E$, and fully determine the symmetric part $\omega_{(a}(\Theta_{b)})$ in terms of $E$ and $T$:
\begin{subeqs}
    \omega_{ab}(V)&= T_{[ab]}-\dd \mathcal{E}_{[ab]}-c\omega_{[a}(\Theta_{b]})
    \\
    \omega_{(a}(\Theta_{b)})&=\frac{1}{c} T_{(ab)}-\frac{1}{c}\dd \mathcal{E}_{(ab)}\\
    \omega_{ab}(\Theta_c)&=\frac{1}{2}\left(\dd \mathcal{E}_{abc}+\dd \mathcal{E}_{bca}-\dd \mathcal{E}_{cab}\right)-\frac{1}{2}\left( T_{abc}+ T_{bca}- T_{cab}\right)
\end{subeqs}
where we have employed the following conventions to simplify the notation
\begin{subeqs}
    T_{ab}&= T_a(V, \Theta_b)  &    T_{abc}&= T_{a}(\Theta_b, \Theta_c)\\
    \dd \mathcal{E}_{ab}&=\dd \mathcal{E}_a(V, \Theta_b) & \dd \mathcal{E}_{abc}&=\dd\mathcal{E}_a(\Theta_b, \Theta_c)\,.
\end{subeqs}
The strongest constraints that can be imposed are $T^a=0$, which lead to the solutions
\begin{eqs}
    \omega_{ab}(\Theta_c)&=\frac{1}{2}\left(\dd \mathcal{E}_{abc}+\dd \mathcal{E}_{bca}-\dd \mathcal{E}_{cab}\right)\\
    \omega_{(a}(\Theta_{b)})&=-\frac{1}{c}\dd \mathcal{E}_{(ab)}\\
    \omega_{ab}(V)&=-\dd \mathcal{E}_{[ab]}-c\omega_{[a}(\Theta_{b]})\,.\label{spin connection}
\end{eqs}
However, the component $\omega_{a}$ remains unconstrained, consistent with the well-known fact that the choice of a non-relativistic connection is not unique. We introduce a vector parameter $\gamma_a$ and an antisymmetric tensor $\gamma_{ab}$ to parameterize these degrees of freedom in the selection of $\omega_{a}$:
\begin{equation}
        \gamma_a\equiv\frac{1}{c}\omega_{a}(V)\qquad\qquad 
    \gamma_{ab}\equiv\frac{1}{c}\omega_{[a}(\Theta_{b]})\,.
\end{equation}

Combining the vielbein postulate \eqref{christoffel} and Eq. \eqref{spin connection} leads to the most general ``torsion-free'' and Galilean-compatible connection coefficients
\begin{eqs}
    \gtGa{^\r_{\m\n}} &=-V^\r\p_\m T_\n+\frac{1}{2}\Pi^{\r\la}\sb{\p_\m\Pi_{\n\la}+\p_\n\Pi_{\m\la}-\p_\la\Pi_{\m\n}}+ \ts{\gamma}{^\r_\m_\n}\label{coe gali}
\end{eqs}
Here
\begin{eqs}
    \ts{\gamma}{^\r_\m_\n}=-T_\m T_\n\tbth{^a^{\r}}\gamma_{a}+\tbth{^{a\r}}\sb{T_\m\tcE{^b_\n}+T_\n\tcE{^b_\m}}\gamma_{ab}\,.
\end{eqs}and $\g_a, \gamma_{[ab]}=\gamma_{ab}$ are arbitrary functions depending on the choice of connection. The expression \eqref{coe gali} agrees with results obtained via second-order computation~\cite{Hartong:2022lsy}. It is worth noting that the ambiguity term is traceless in its first two indices due to the antisymmetry of $\gamma_{ab}$
\begin{eqs}
     \ts{\gamma}{^\r_\r_\n}=0+T_\n \delta^{ab}\gamma_{ab}=0
\end{eqs}which is the direct consequence of partly torsionless condition.

The torsion associated with $\gtGa{^\r_{\m\n}}$ is given by
\begin{eqs}
    \gtGa{^\r_{[\m\n]}}=-V^\r \sb{\dd T}_{\m\n}\,.
\end{eqs}

Note that the action of the pre-Galilean covariant derivative on $V^\m$ and $\Pi_{\m\n}$ yields two non-vanishing covariant expressions
\begin{eqs}
     \gtna{_\m}V^\n&=-\Pi^{\n\s}\mathcal{K}_{\s\m}+\ts{\gamma}{^\n_\m_\r}V^\r\\
       \gtna{_\r}\Pi_{\m\n}&=-T_{(\m}\mathcal{K}_{\n)\r}-2\ts{\gamma}{^\s_\r_{(\m}}\Pi_{\n)\s}
\end{eqs}
where $\mathcal{K}_{\m\n}=-\frac{1}{2}\mathcal{L}_V \Pi_{\m\n}$ is identified as the extrinsic curvature.

The trace of the pre-Galilean connection coefficients is
\begin{eqs}
    \gtGa{^\r_{\r\n}}&=-V^\r\p_\r T_\n+\delta_{ab}\tcE{^a^\r}\p_{\n}\tcE{^b_\r}=-\mathcal{L}_V T_\n +\frac{1}{E}\p_{\n} E
\end{eqs}
which leads to a useful expression for the covariant divergence of an arbitrary vector field $\mathbf{V}^\m$
\begin{eqs}
    \gtna{_\m}\mathbf{V}^\m=\frac{1}{E}\p_\m\sb{E\mathbf{V}^\m}-\mathbf{V}^\m A_\m\,.\label{g div}
\end{eqs}

\subsection[Large-\texorpdfstring{$c$}{c} expansion of Ricci scalar]{Large-\texorpdfstring{$\boldsymbol{c}$}{c} expansion of the Ricci scalar in second-order variables}
We now proceed to expand the Ricci scalar in powers of $c$ by utilizing the $(1+d)$ metric decomposition defined in Eqs. \eqref{1+d a} and \eqref{1+d b}. We first express the Lorentzian Ricci tensor in terms of the pre-Galilean variables.

The first step involves decomposing the Levi-Civita connection into terms of varying orders in $c$
\begin{eqs}
    \lctGa{^\r_\m_\n}&=\frac{1}{c^2}\mttGa{^\r_\m_\n}+\gtGa{^\r_\m_\n}+\ts{S}{^\r_\m_\n}-\ts{\gamma}{^\r_\m_\n}+\ttGa{^\r_\m_\n}
\end{eqs}
where the constituent tensors are defined as
\begin{subeqs}
    \mttGa{^\r_\m_\n}&=-V^\r\mathcal{K}_{\m\n}\\
    \ts{S}{^\r_\m_\n}&=-V^\r T_{(\m}A_{\n)}+\frac{1}{2}V^\r T_{\m\n}\\
    \ttGa{^\r_\m_\n}&=\Pi^{\r\la}T_{\la(\m}T_{\n)}
\end{subeqs}with $T_{\m\n}=\sb{\dd T}_{\m\n}$ and $A_\m=\mathcal{L}_V T_\m$.

We use the convention that the derivative index is the first lower index of the connection,
\begin{eqs}
    \nabla_\s \tT{^\r_\m_\n}
    =\p_\s\tT{^\r_\m_\n}
    +\tGa{^\r_\s_\la}\tT{^\la_\m_\n}
    -\tGa{^\la_\s_\m}\tT{^\r_\la_\n}
    -\tGa{^\la_\s_\n}\tT{^\r_\m_\la}\,
\end{eqs}
With $\tT{^\r_\m_\n}=2\tGa{^\r_{[\m\n]}}$, our curvature convention is
\begin{eqs}
    [\nabla_\m,\nabla_\n]w_\s
    =\tR{_\m_\n_\s^\r}w_\r-\tT{^\r_\m_\n}\nabla_\r w_\s\,
\end{eqs}
where
\begin{align}
    \tR{_\m_\n_\s^\r}
    =-\p_\m\tGa{^\r_\n_\s}+\p_\n\tGa{^\r_\m_\s}
    +\tGa{^\la_\m_\s}\tGa{^\r_\n_\la}
    -\tGa{^\la_\n_\s}\tGa{^\r_\m_\la}\,
    \label{Riemann D}
\end{align}
The Ricci tensor is obtained by contraction,
\begin{align}
    R_{\m\n}:=\tR{_\m_\r_\n^\r}
    =-\p_\m\tGa{^\r_\r_\n}+\p_\r\tGa{^\r_\m_\n}
    +\tGa{^\la_\m_\n}\tGa{^\r_\r_\la}
    -\tGa{^\la_\r_\n}\tGa{^\r_\m_\la}\,
    \label{Ricci def}
\end{align}

Expanding the Ricci tensor in powers of $c$ and organizing the terms order by order in a covariant form yields
\begin{align}
    R_{\m\n}=\frac{1}{c^4}\mftR{_\m_\n}+\frac{1}{c^2}\mttR{_\m_\n}+\ztR{_\m_\n}+c^2\ttR{_\m_\n}+c^4\ftR{_\m_\n}\,.
\end{align}
Here, the expansion coefficients are given by
\begin{subeqs}
    \mftR{_\m_\n}&=0\\
    \mttR{_\m_\n}&=\gtna{_\r} \mttGa{^{\r}_{\m\n}}-2\ts{\gamma}{^{\la}_{\r(\m}}\mttGa{^{\r}_{\n)\la}}\\
  \ztR{_{\mu\nu}} &= \gtR{_{\mu\nu}} + \gtna{_{\rho}} \ts{S}{^{\rho}_{\mu\nu}} - \gtna{_{\mu}} \ts{S}{^{\rho}_{\rho\nu}} -\ts{S}{^\r_\la_\m}\ts{S}{^\la_\r_\n} - 2\mttGa{^{\lambda}_{\rho(\mu}} \ttGa{^{\rho}_{\nu)\lambda}}\notag\\
&\quad\quad\quad\quad\quad\quad\quad\gtna{_\r}\ts{\gamma}{^\r_\m_\n} -\ts{\gamma}{^\r_\m_\n}\ts{S}{^\s_\s_\r}+2\ts{S}{^\r_\la_(_\m}\ts{\gamma}{^\la_\n_)_\r}-\ts{\gamma}{^\r_\m_\la}\ts{\gamma}{^\la_\r_\n}\\
    \ttR{_{\m\n}}&=\gtna{_\r} \ttGa{^{\r}_{\m\n}}+\ttGa{^{\la}_{\m\n}} \ts{S}{^{\r}_{\r\la}} -2\ts{S}{^\r_\la_(_\m}\ttGa{^\la_\n_)_\r}+2\ts{\gamma}{^\r_\la_(_\m}\ttGa{^\la_\n_)_\r}\\
    \ftR{_\m_\n}&=-\frac{1}{4}\Pi^{\m\r}\Pi^{\n\s}T_{\m\n}T_{\r\s}\,.
\end{subeqs}
We now determine the $c$-expansion of the Ricci scalar
\begin{eqs}
    R&=\frac{1}{c^4}\mftR{}+\frac{1}{c^2}\mttR{}+\ztR{}+c^2\ttR{}+c^4\ftR{}\,.\\
\end{eqs}To simplify these expressions, we employ the following useful identities
\begin{subeqs}
    \ts{S}{^\r_\m_\n}V^\n&=0&\quad \ts{S}{^\r_\m_\n}V^\m&=V^\r A_\n\\
    \ts{S}{^\r_\r_\m}&=A_\m&\quad \ts{S}{^\r_\m_\r}&=0\\
    \tbth{_a^\m}\tbth{_b^\n}\ts{\gamma}{^\r_\m_\n}&=0&\quad V^\m V^\n\ts{\gamma}{^\r_\m_\n}&=-\gamma^a\tbth{_a^\r}
\end{subeqs}
The non-vanishing terms in this expansion are found to be
\begin{subeqs}
    \mttR{}&=-V^\m V^\n \gtR{_\m_\n}+\gtna{_\r}\sb{\gamma^a\tbth{_a^\r}}-\gamma^a\tbth{_a^\m}A_\m-3\gamma^{ab}\gamma_{ab}-\gtna{_\r}\sb{V^\r \mathcal{K}}\\
    \ztR{}&=\Big(-\gtna{_\m} A^\m-A^\m A_\m\Big) +\Big(\Pi^{\m\n}\gtR{_\m_\n}-\gtna{_\m}A^\m-A^\m A_\m-2\gamma^{ab}\tbth{_a^\m}\tbth{_b^\n}T_{\m\n}\Big)\\
        \ttR{}&=\frac{1}{4}\Pi^{\m\n}\Pi^{\r\s}T_{\m\r}T_{\n\s}\,.
\end{subeqs}

Finally, we focus on 2d spacetime. In the $d=1$ case, $\gamma_{ab}=0$, and the terms $\Pi^{\m\n}\gtR{_\m_\n}$ and $\ttR{}$ vanish, since $\gtna{_\r}\tbth{^\m}=0$.
Consequently, the Ricci scalar in $(1+1)$ dimensions reduces to
\begin{eqs}
    R^{(2d)}&=\frac{1}{c^2}\big(-V^\m V^\n\gtR{_\m_\n}+\gtna{_\m}\sb{\gamma \tbth{^\m}}-\gtna{_\m}\sb{V^\m \mathcal{K}}-\gamma \tbth{^\m} A_\m\big)-2\big(\gtna{_\m}A^\m+A^\m A_\m\big).\label{c expansion of Ricci}
\end{eqs}



\bibliographystyle{fullsort}
\bibliography{ref}

\end{document}